\documentclass{aa}  
\usepackage{graphics,graphicx}
\usepackage{color}
\usepackage{epsf}
\usepackage{epsfig}
\usepackage{dcolumn}

\newcommand{\ka}{\mbox{WR\,102ka}}
\newcommand{\wrc}{\mbox{WR\,102c}}

\newcommand{\Lsun}{\mbox{$L_\odot$}}
\newcommand{\Msun}{\mbox{$M_\odot$}}
\newcommand{\vinf}{\mbox{$\varv_\infty$}}
\newcommand{\mdot}{\mbox{$\dot{M}$}}
\newcommand{\myr}{\mbox{$M_\odot\,{\rm yr}^{-1}$}}
\newcommand{\mim}{\mbox{$\mu$m}}

\newcommand{\lsim}{\raisebox{-.4ex}{$\stackrel{<}{\scriptstyle \sim}$}}
\newcommand{\msim}{\raisebox{-.4ex}{$\stackrel{>}{\scriptstyle \sim}$}}

\def \etal   {\hbox{et~al.\/}}

\def\changed{}
\usepackage{txfonts}
\begin{document}
\titlerunning{Two extremely luminous WN stars in the Galatic Center}
\title{Two extremely luminous WN stars in the Galactic center with 
circumstellar emission from dust and gas}

\author{A. Barniske, L.\,M. Oskinova, and W.-R. Hamann
          }
          
\offprints{lida@astro.physik.uni-potsdam.de}

\institute{Potsdam University, Am Neuen Palais 10, Potsdam, 14469 Germany
}

\date{Received 12 February 2008 / Accepted  22 May 2008}

% \abstract{}{}{}{}{} 
% 5 {} token are mandatory
\abstract
% Context
{The central region of our Galaxy contains a large population of young
massive stars. These stars are concentrated in three large star
clusters, as well as being scattered in the field. Strong ionizing radiation
and stellar winds of massive stars are the essential feedback agents that 
determine the physics of the ISM in the Galactic center.}
{The aim is to study relatively isolated massive WN-type stars in the
Galactic center in order to explore their properties and their
influence on the ISM. }
{The K-band spectra of two WN stars in the Galactic center, \ka\ and
\wrc, are exploited to infer the stellar parameters and 
to compute synthetic stellar spectra using the Potsdam Wolf-Rayet
(PoWR) model atmosphere code. These models are combined with
dust-shell models for analyzing the {{\em Spitzer}} IRS spectra of
these objects. Archival IR images complement the interpretation.  }
{We report that \ka\ and \wrc\ are among the most luminous stars in
the Milky Way. They critically influence their immediate environment
by strong mass loss and intense UV radiation, and thus set the
physical conditions for their compact circumstellar nebula.  The mid-IR
continua for both objects are dominated by dust emission.  For the
first time we report the presence of dust in the close vicinity of WN
stars. Also for the first time, we have detected lines of pure-rotational
transitions of molecular hydrogen in a massive-star nebula.  A
peony-shaped nebula around \ka\ is resolved at 24\mim\ by the {{\em
Spitzer}} MIPS camera. We attribute the formation of this IR-bright 
nebula to the recent evolutionary history of \ka.  }
{}


\authorrunning{Barniske, Oskinova, Hamann }
 \maketitle
%
%________________________________________________________________

\section{Introduction}

Visual light cannot penetrate the dust clouds obscuring the innermost
part of our Galaxy, whereas infrared astronomy opens extraordinary
views on this environment.   Three very massive star clusters have been
discovered in the Galactic center (GC) during the last decade.  The
central cluster is located within 1\,pc from the central black hole
(Krabbe \etal\ \cite{krab95}). Two other massive star clusters, the
Arches and the Quintuplet, are located within 30\,pc projected distance
from Sgr\,A* (Serabyn \etal\ \cite{arch98}, Figer \etal\
\cite{quin99}).  While the Arches cluster is younger and contains many
OB-type stars, the more evolved Quintuplet cluster (3-5\,Myr old)
harbors many Wolf-Rayet (WR) stars. Besides these compact
stellar conglomerates, many high-mass stars whose association with 
stellar clusters is not obvious are scattered in the GC. Among these
are the rather isolated WR type stars \ka\ and \wrc\ -- our
program stars.

Massive stars severely influence their environment by strong ionizing
radiation, mass and kinetic energy input (Freyer \etal\ \cite{fr03}).
The thermal Arched Filaments in the vicinity of the Arches and the  
Sickle nebula in the vicinity of the Quintuplet are thought to be
powered by the combined action of hot massive stars located in these
clusters (Simpson \etal\ \cite{sim07} and references therein). However,   
radio (e.g. Yusef-Zadeh \& Morris \cite{yz87}) and infrared (Price \etal\
\cite{msx01}, Rodr{\'i}guez-Fern{\'a}ndez \etal\ \cite{rf01},
Simpson \etal\ \cite{sim07}) surveys of the GC reveal a
complex structure of the ionized gas with many small-scale compact
sources of thermal emission. In this paper we investigate the emission
from such nebulae around our program stars.

A star with initial mass over $\approx$20\,\Msun\ settles on the
main sequence as an O-type. Stars more massive than
$\approx$30\,\Msun\ evolve off the main sequence at more or less
constant luminosity, but their mass-loss rate increases significantly.
Stars with initial masses above $\approx$40\,\Msun\ pass through a
short ($\sim10^5$\,yr) luminous blue variable (LBV) evolutionary stage
that is characterized by high mass-loss rates and violent eruptions.
The ejected material, up to several $M_\odot$, is often observed in
the form of an associated nebula -- one famous example is the
Homunculus nebula around the LBV star $\eta$\,Car.  The most massive
stars with initial masses $M_{\rm i} \msim$90\,\Msun\ are thought to
lose enough mass during their life on the main-sequence to evolve to
the WR stage without ever becoming an LBV.

Stars that display CNO-processed matter in a strong stellar wind are
classified as WR stars of the nitrogen sequence (WN type).  The cooler,
late WN subtypes (WNL) usually contain some rest of hydrogen in their
atmospheres, while the hotter, early subtypes (WNE) are hydrogen free
(Hamann \etal\ \cite{ham91}).  {\changed The WNL evolutionary stage can,
in fact, precede the LBV stage (Langer et al.\ \cite{lan94}, Smith \&
Conti \cite{smith08}). Typically, WNL stars are significantly more
luminous than WNE stars (Hamann \etal\, \cite{wrh06}).} The WN phase
may be followed by the WC stage, when the products of helium burning
appear in the stellar atmosphere.  Wolf-Rayet stars end their lives with
a super- or hypernova explosion.

During its evolution a massive star loses a considerable fraction of
its initial mass. This material accumulates in the circumstellar
environment (e.g.\ Freyer \etal\ \cite{fr03}).  Recently, concerns
were raised that the empirically derived stellar mass-loss rates need
to be drastically reduced (Fullerton
\etal\ \cite{ful06}).  However, when the inhomogeneous 
nature of stellar winds is accounted for, the analyses of optical, UV,
and X-ray spectra of massive stars consistently yield mass-loss rates
that are only factor of two lower than inferred under the
``standard`` assumption of smooth winds (Oskinova \etal\ \cite{osk07}).

Infrared (IR) spectra of H\,{\sc ii} regions around main-sequence stars are
composed of nebular emission lines and a dust-dominated continuum
(Dopita \etal\ \cite{dop05}). However, in the vicinity of WR stars
dust is rarely found, except for WC-type stars in the inner,
metal-rich parts of galaxies (e.g.\ Crowther \etal\, \cite{wc9}).
These stars are surrounded by dust shells, and it appears that the
metal-rich environment of the GC is 
favorable to the formation of
circumstellar dust. For WN stars, however, it is generally believed
that the chemical composition and strong radiation prohibit the
formation of dust in their vicinity.

%%%%%%%%%%%%%%%%%%%%%%%%%  FIG 1  %%%%%%%%%%%%%%%%%%%%%%%%%%%%%%%%
%-------------------------------------------------------------
\begin{figure}
  \centering \includegraphics[width=\columnwidth]{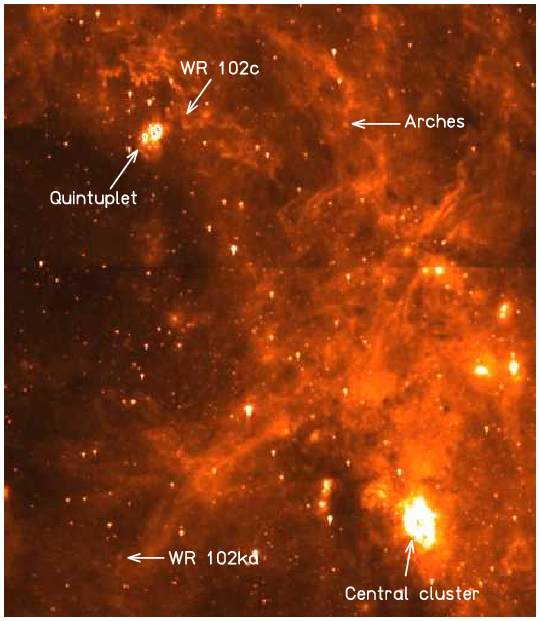}
  \caption{{\em Spitzer} IRAC 8\,\mim\ archive image of the GC.  
  Arrows point to the locations of our program stars, the Arches, 
  the Quintuplet, and the Central clusters. North is to the top and 
  east to the left. }  \label{fig:gc_reg}
\end{figure}
%
%_____________________________________________________________

We secured the mid-IR spectra of WN-type stars \ka\ and \wrc\ with the
{\em Spitzer} Space Telescope. Our program stars are sufficiently
isolated to allow high-resolution spectroscopy with the
{\it Spitzer} infrared spectrograph IRS. The basic strategy for the
analysis is to model the stellar spectrum using a stellar atmosphere
code, and then use the synthetic stellar spectrum as input for
modeling the circumstellar nebula.

This paper is organized as follows. Stellar parameters and synthetic
stellar spectra of our program stars are obtained in Section\,2 from
the analysis of K-band spectra. Infrared spectroscopic and imaging
observations are presented in Sect.\,3. The analysis and modeling of the
mid-IR {\em Spitzer} IRS spectra are conducted in
Sect.\,4. A discussion is presented in Sect.\,5, followed by the
summary in Sect.\,6.

%%%%%%%%%%%%%%%%%%%%%%%%%%%%%%%%%%%%%%%%%%%%%%%%%%%%%%%%%%%%%%%%%%%%%%%%%%%%
%--------------------  Section 1 PoWR ------------------------------
%%%%%%%%%%%%%%%%%%%%%%%%%%%%%%%%%%%%%%%%%%%%%%%%%%%%%%%%%%%%%%%%%%%%%%%%%%%%%
%%%%%%%%%%%%%%%%%%%%%%%%%%%%%%%%%%%%%%%%%%%%%%%%%%%%%%%%%%%%%%%%%%%%%%%%%%%%%

\section{Program stars \ka\ and \wrc}
\label{sect:analysis}

%%%%%%%%%%%%%%%%%%%%%%%%%  FIG 2 %%%%%%%%%%%%%%%%%%%%%%%%%%%%%%%%
%-------------------------------------------------------------
\begin{figure}
  \centering
  \includegraphics[width=\columnwidth]{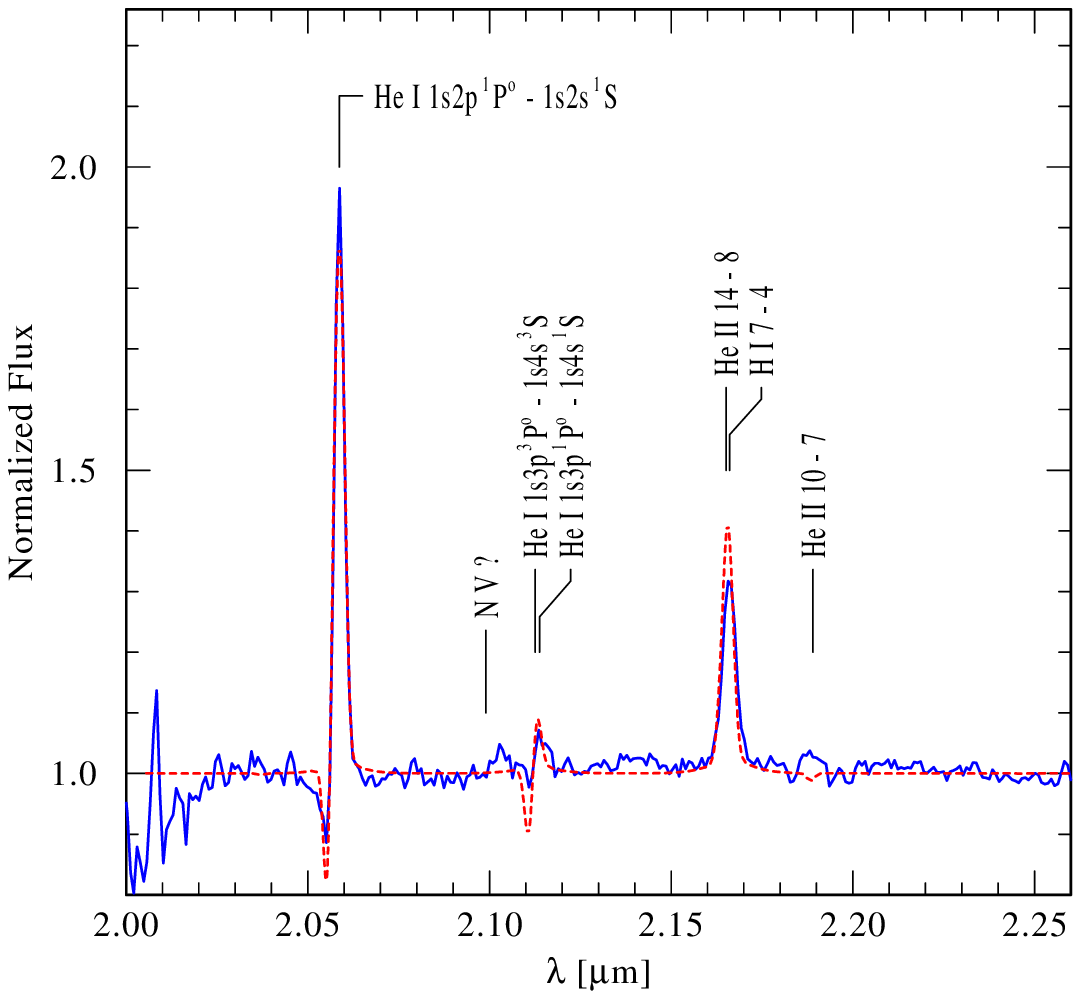}
  \caption{K-band spectrum of \ka\ (blue solid line) together with 
    the best-fitting PoWR model spectrum (red dashed line)}
  \label{fig:k102ka}
\end{figure}
%
%_____________________________________________________________

Figure\,\ref{fig:gc_reg} shows the location of our program stars relative to
the three massive clusters in the GC -- the Central Cluster, the Arches
cluster, and the Quintuplet cluster.

%--------------------------------------------------------------------
\begin{table}
\caption{Measured equivalent widths  in the K-band spectral lines of
\ka } 
\label{tab:ka_k}
\centering
\begin{tabular}{lcD{.}{.}{-1}}
%\begin{tabular}{llr}
\hline
\hline
\rule[0mm]{0mm}{3.25mm}Line & $\lambda$ & \multicolumn{1}{c}{$-W_\lambda$} \\
                            & [\mim ]   & \multicolumn{1}{c}{[\AA]}     \\
\hline
\rule[0mm]{0mm}{3.25mm}He\,{\sc i}  & 2.059 & 26.3   \\
He\,{\sc i}                         & 2.112 & 2.3    \\
He\,{\sc ii}+Br$\gamma$             & 2.165 & 15.4   \\
He\,{\sc ii} (?)                    & 2.189 & <2.0   \\ 
\hline
\end{tabular}
\end{table}
%------------------------------------------------------------------------

\object{WR\,102ka} has a projected distance of 19\,pc from Sgr\,A$^\ast$  
and apparently does not belong to any star cluster.  It was first
observed during a near-infrared survey in the GC by Homeier \etal\
(\cite{H03}). By comparing the K-band spectrum of
\ka\ with the spectra of two WR stars in the LMC, \ka\ was classified
as a WN10 spectral subtype.

%%%%%%%%%%%%%%%%%%%%%%%%%  FIG 3 %%%%%%%%%%%%%%%%%%%%%%%%%%%%%%%%
%-------------------------------------------------------------
%
\begin{figure*}
\centering
\includegraphics[width=1.95\columnwidth]{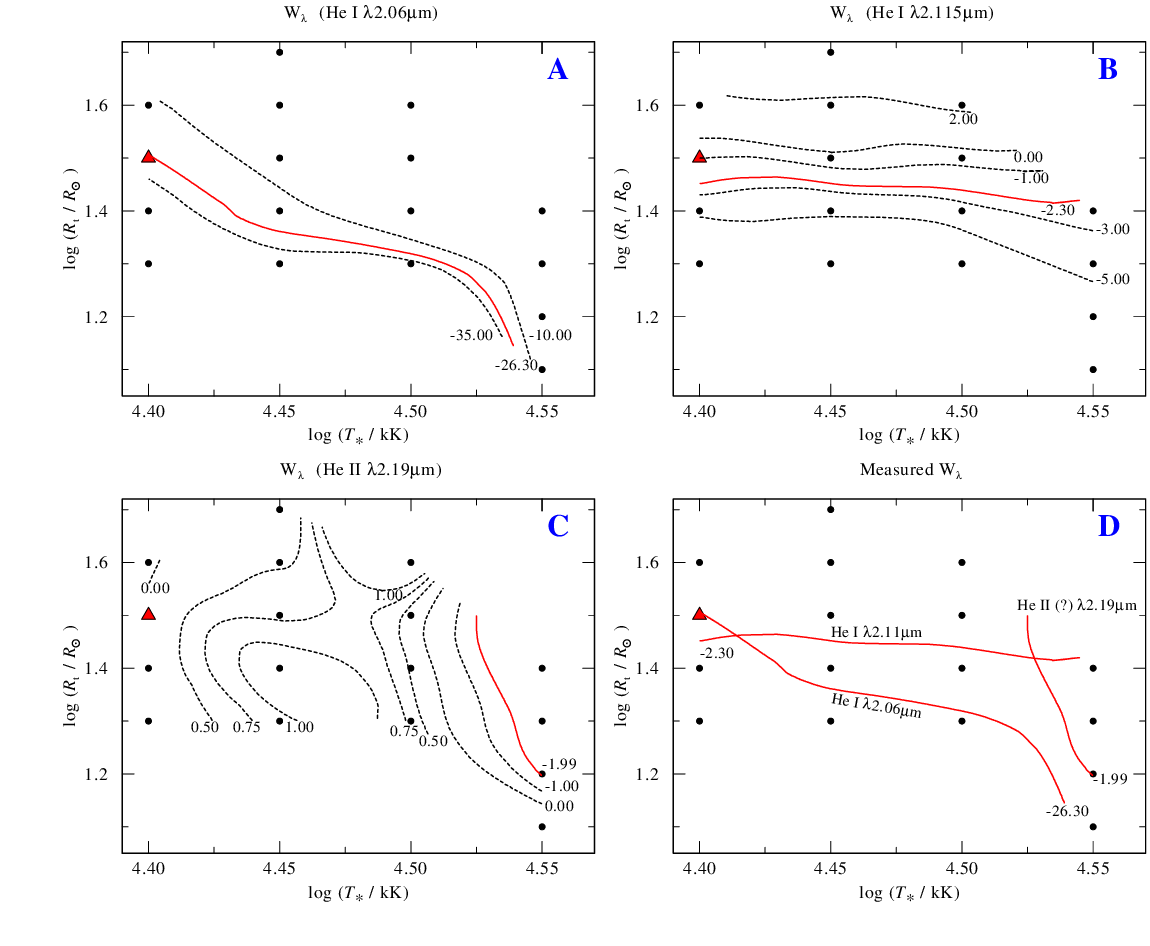}
\caption{Contours of constant line emission of helium in the
$\log{T_\ast}\,-\,\log{R_{\rm t}}$ plane. All other parameters are kept
constant ($\log L_\ast/\rm{\Lsun} = 6.3$, $\varv_\infty = 400$\,km/s).
Dots indicate the calculated grid models. The interpolated contour
lines are labeled with the equivalent width in \AA. The solid red lines
correspond to the measured value of $W_{\lambda}$.
Panel {\bf A}: He\,{\sc i} $\lambda\,2.06$\,\mim; 
{\bf B}: He\,{\sc i} $\lambda\,2.115$\,\mim; 
{\bf C}: He\,{\sc ii} (?) $\lambda\,2.19$\,\mim. 
The contours corresponding to the measured $W_{\lambda}$ are shown 
together in panel {\bf D}. The large triangle indicates the adopted final
model for \ka\ (see text).}
\label{fig:isocont}
\end{figure*}
%
%_____________________________________________________________

\object{WR\,102c} lies in the outskirts of the Quintuplet cluster, in a 
large arc of diffuse emission called the Sickle nebula. The star was
discovered during a survey by Figer \etal\ (\cite{Fig99}), and
classified as WN6 subtype because the similarity of its K-band
spectrum with the well-studied WN6 type star WR\,115. 
{\changed It should be noted that the WN6 classification criteria can be 
met by either hydrogen-free WNE subtypes or by WNL stars with hydrogen. 
As \wrc\ shows hydrogen in its spectrum (cf.\ Sect.\,2.2), it belongs to 
the WNL subclass.}  

The coordinates of WR\,102ka from the discovery paper are in agreement
with the \textrm{2MASS} point source catalog.  For WR\,102c, however,
we report a difference of 0$\fs$3 in RA between the coordinates given
in Figer \etal\ (\cite{Fig99}) and the 2MASS point source catalog. For
our {\it Spitzer} observation we used the latter coordinates (cf.\
Table\,\ref{table:obs}). {\changed These coordinates agree
with the catalog of point sources from {\it Spitzer} IRAC
observations of the central part of the Galaxy (Ram{\'i}rez \etal\ \cite{ram08}).}
Throughout this paper we assume the distance to our program stars as 
$d=8$\,kpc (Reid \cite{r93}).

\subsection{Stellar parameters of \ka}

The only available part of the stellar spectrum of \ka\ is the near-IR
K-band spectrum obtained with SOFI at the ESO 3.6\,m New Technology
Telescope (Homeier \etal\ \cite{H03}). As can be seen in
Fig.\,\ref{fig:k102ka}, the spectrum is dominated by strong emission
lines of He\,{\sc i} and He\,{\sc ii}\,+\,Br$\gamma$. A small, flat-topped
emission feature is present at the wavelength of the He\,{\sc
ii}$\,10-7$ transition. However, this feature appears broader than the
other lines and has a different spectral shape. Neither nebular 
(forbidden) lines nor H$_2$ fluorescent emission is visible  
in the K-band spectrum of \ka. The measured equivalent widths 
are listed in Table\,\ref{tab:ka_k}.

The Potsdam Wolf-Rayet (PoWR) stellar atmosphere models are employed for
the analysis of the K-band spectrum. The PoWR code solves the non-LTE
radiative transfer in a spherically expanding atmosphere, consistently
with the statistical equations and energy conservation. Iron-line
blanketing and wind clumping in first approximation are taken into 
account (Hamann \etal\ \cite{wrh04}). 
Grids of models for WN stars can be found
on the Potsdam Wolf-Rayet (PoWR) models web-site
\footnote{\hbox{http://www.astro.physik.uni-potsdam.de/PoWR.html}}. 

All lines that can be definitely identified in the K-band spectrum of \ka\ 
are due to helium and hydrogen; therefore
we cannot determine other element abundances. We adopt mass fractions
that are typical for Galactic WN stars -- N:\,0.015, C:\,0.0001,
Fe:\,0.0014 (Hamann \etal\ \cite{wrh04}) -- throughout this paper.

The terminal wind velocity in \ka, $\vinf\approx 400$\,km s$^{-1}$, is
inferred from fitting the He\,{\sc i} line profiles. To assess the
hydrogen abundance, we use the line blend He\,{\sc ii}\,+\,Br$\gamma$.
Our models show that whenever the strong observed He\,{\sc i} lines
are reproduced, the He\,{\sc ii} emissions are very small. Thus the
strong emission at 2.165\,\mim\ must be mainly due to hydrogen; we
obtain a hydrogen mass fraction of 0.2 for \ka, which is typical for
late-type WN (WNL) stars.

For a given chemical composition and stellar temperature $T_\ast$,
synthetic spectra from WR model atmospheres of different mass-loss
rates, stellar radii and terminal wind velocities yield almost the same
emission line equivalent widths, if they agree in their "transformed
radius" $R_{\rm t}$ defined as 
\begin{equation}
R_{\rm t} = R_*
  \left[\frac{\varv_\infty}{2500 \, {\rm km}\,{\rm s^{-1}}} \left/
  \frac{\sqrt{D} \dot M}
       {10^{-4} \, M_\odot \, {\rm yr^{-1}}}\right]^{2/3} \right .
\end{equation}
For the clumping contrast we adopt $D=4$ as a typical value for WN
stars (Hamann \etal\ \cite{hk98}). Note that $R_{\rm t}$ is inversely 
correlated with the mass-loss rate, i.e.\ the smaller the transformed 
radius the higher is the density in the stellar wind.

In order to derive the fundamental stellar parameters of \ka\ in a
systematic way, we 1) calculate a reasonably fine grid of models in the
adequate parameter range;  2) evaluate the equivalent widths
($W_\lambda$) of model lines; 3) compare modeled $W_\lambda$ with
measured ones and choose the model which is capable to simultaneously
reproduce the measured equivalent widths.

Contours of constant line emission in the model grid are shown in
Fig.\,\ref{fig:isocont}.  Figure\,\ref{fig:isocont} shows a remarkable
difference between the two He\,{\sc i} lines in the K-band (compare
Panels\,A and B). While the He\,{\sc i}\,$\lambda 2.06$\,\mim\ singlet
line is sensitive to both model parameters $R_{\rm t}$ and $T_\ast$,
the singlet/triplet blend He\,{\sc i}\,$\lambda 2.115$\,\mim\ has a
$W_\lambda$ which is nearly independent of $T_\ast$.

Fig.\,\ref{fig:isocont}D reveals that no consistent fit is possible
for the three considered helium lines. We attribute the problem to the
line at 2.189\,\mim, tentatively identified with the He\,{\sc ii}\,10-7
transition. Considering that the small observed $\lambda\,2.189$\,\mim\
feature is broader than other lines, that it has a different line profile,
and that the available spectrum is quite noisy, we conclude that the
$\lambda\,2.189$\,\mim\ feature cannot be due to He\,{\sc ii} 
emission from that star.

Therefore we chose as our best-fitting model the grid point that lies
closest to the intersection point of the contours for the strong He\,{\sc
i}\,$\lambda 2.06$\,\mim\ and He\,{\sc i}\,$\lambda 2.115$\,\mim\ lines. 
The parameters are $T_*$ = 25.1\,kK and $\log R_{\rm t}=1.48$  (large 
triangle in the panels of Fig.\,\ref{fig:isocont}).

%--------------------------------------------------------------------
\begin{table}
\caption{Stellar parameters of \ka\ and \wrc} 
\label{tab:par}
\centering
\begin{tabular}{lll}
\hline
\hline %---------------------------------------------------------------
\rule[0mm]{0mm}{3.25mm}                     & \ka            & \wrc \\
\hline %---------------------------------------------------------------
\rule[0mm]{0mm}{3.25mm}Spectral type        & Ofpe/WN9       & WN6(h?)\\  
\vinf\ [km\,s$^{-1}$]                       & 400            & $\ge 1300$\\
%$\log\,L\ [\Lsun]$                      & $6.5 \pm 0.2$  & $6.3 \pm 0.4$\\
$\log\,L\ [\Lsun]$                      & $6.5 \pm 0.2$  & $6.3 \pm 0.3$\\
$^{(a)}\log\,\Phi_{\rm{i}}\ [\rm s^{-1}]$         & $48.96$        & 50.12 \\
%$A_{V}\ [{\rm mag}]$                        & $27\pm 5$      & 25.7 \\    
$A_{V}\ [{\rm mag}]$                        & $27\pm 5$      & $26\pm 1$ \\
%$E_{B\,-\,V}\ [{\rm mag}]$                  & $8.0\pm 1$     & 7.56\\
$E_{B\,-\,V}\ [{\rm mag}]$                  & $8.0\pm 1$     & $7.6\pm 0.3$\\ 
$T_*$\ [kK]                                 & 25.1           & $\approx\,50$ \\
$\log\,\dot{M}\ [\myr]$                     & $-4.4$         & $-4.0$ \\
$R_\ast\ [\rm{R_{\odot}}$]                  & 92             & 20\\
\hline %----------------------------------------------------------------
 \multicolumn{3}{l}{(a)\,$\Phi_{\rm{i}}$: number of H ionizing photons per second}
\end{tabular}
\end{table}
%------------------------------------------------------------------------

After we have derived $R_{\rm{t}}$ and $T_\ast$ from the normalized line
spectrum, the absolute values of $L$, $R_\ast$ and \mdot\ are obtained by
fitting the spectral energy distribution (SED). The absolute flux scales
proportional to $R_\ast^2$ which in turn means that for a fixed value of
$R_{\rm t}$ the mass loss rate \mdot\ is proportional to $L^{3/4}$. For
convenience we calculate our models with a 'generic' luminosity of
$\log\,{L/\Lsun} = 6.3$ and scale it to match the observations.
$R_\ast$ and \mdot\ are then scaled along with the luminosity using the
relations mentioned above.

To account for the interstellar absorption we adopt the reddening law
by Moneti \etal\ (\cite{m2001}), which was obtained for the Quintuplet
cluster region and is therefore suitable for our program stars. In
this law, the ratio between the V- and K-band extinctions is $A_V/A_K = 8.9$.

\begin{figure}
\centering
\includegraphics[width=0.9\columnwidth]{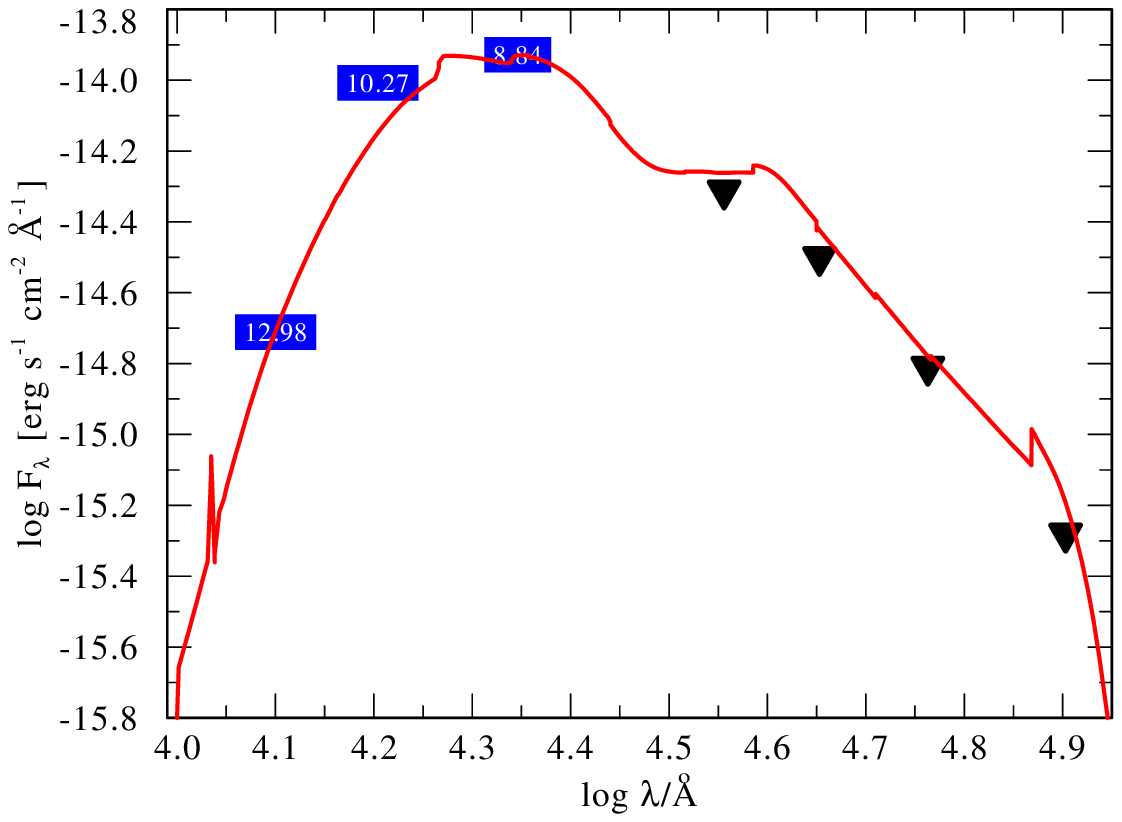}
\caption{Spectral energy distribution for \ka. The thick red line gives the
reddened model flux (see Table\,\ref{tab:par} for the parameters). Boxes give
observed 2MASS magnitudes (labels). Triangles correspond to the flux in
the four IRAC channels, which we extracted from the archival images (see
Sect.\,\ref{sec:irac} and Table\,\ref{tab:irac}).}
\label{fig:phka}
\end{figure}

By scaling the luminosity (logarithmic shift) and simultaneously
varying $E_{B-V}$, we adjust the synthetic SED to the 2MASS and the
{\em Spitzer} IRAC photometry marks (see Sect.\,\ref{sec:irac}). The
best fit is obtained with $E_{B-V} = 8.0 \pm 1$\,mag and is shown in
Fig.\,\ref{fig:phka}. The stellar parameters of \ka\ are compiled in
Table\,\ref{tab:par}. With a bolometric luminosity of $\log\,L/\Lsun =
6.5\,\pm\,0.2$, \ka\ is one of the most luminous stars in the Galaxy!

According to its location in the $T_\ast-R_{\rm t}$ plane
(cf.\ Hamann \etal\ \cite{wrh04}), \ka\ has a spectral type later than 
WN8. The mass-loss rate and wind velocity of
\ka\ found from our modeling are very similar to the parameters of
Ofpe/WN9 stars in the GC as determined by Martins
\etal\,(\cite{mar07}). Their Ofpe/WN9 objects cover a temperature range
form $T_\ast = 20$\,kK ... 23\,kK, terminal velocities lie between 450
and 700\,km/s, and the mass-loss rates are 
$\log\,\mdot\,=\,-4.95\,...\,\mbox{-4.65}$. Hence we adopt the spectral 
classification
Ofpe/WN9 for \ka.

%%%%%%%%%%%%%%%%%%%%%%%%%  FIG 4 %%%%%%%%%%%%%%%%%%%%%%%%%%%%%%%%
%-------------------------------------------------------------
%
%_____________________________________________________________

%%%%%%%%%%%%%%%%%%%%%%%%%  FIG 5  %%%%%%%%%%%%%%%%%%%%%%%%%%%%%%%%
%------------------------------------------------------------
\begin{figure}
 \centering \includegraphics[width=\columnwidth]{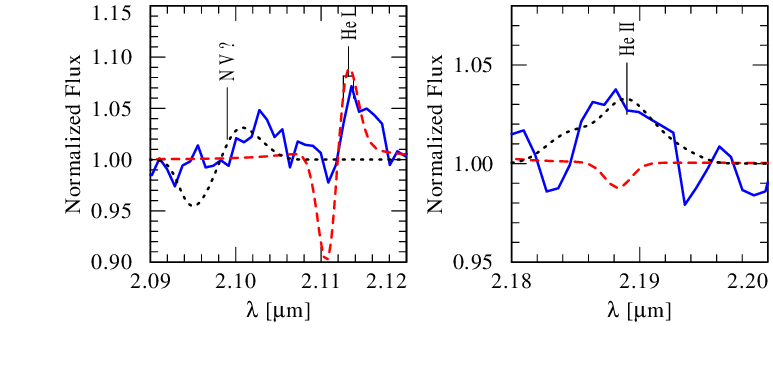}
\caption{Features in the  \ka\  spectrum that indicate the possible 
presence of a companion. Observations (blue ragged line) are compared
to the synthetic spectrum for our final model of \ka\ (red dashed
line).  The black dotted line represents the spectrum of a hot WNE
model, scaled down by a factor of ten.  By superimposing both models,
the N\,{\sc v}\,(?), He\,{\sc i}, and He\,{\sc ii} lines can be reproduced
simultaneously.}
\label{fig:nv}
\end{figure}
%_____________________________________________________________

As discussed above, our final model for \ka\ cannot reproduce the weak
emission feature at $\lambda\,2.189$\,\mim. Instead, a small
absorption feature from He\,{\sc ii} is predicted at this
wavelength. Moreover, there is a weak emission feature around
2.103\,\mim\ visible in the observed spectrum of \ka, which could be
the emission wing of a P-Cygni profile from N\,{\sc v}
$\lambda$2.099\,\mim; however, the appearance of this line is also not
predicted by our final model.  Therefore we must consider the
possibility that \ka\ is actually a binary system, and the spectrum is
contaminated by a hotter but fainter companion. The N\,{\sc v} lines
at 2.099\,\mim\ appears in models for hot WNE stars, as demonstrated
in Fig.\,\ref{fig:nv} by adding a spectrum from a corresponding model
($T_\ast =$ 178\,kK, $\log R_{\rm t} =$ 0.8). The presence of a
WNE-type companion with fast stellar wind could also explain the
appearance of the He\,{\sc ii} $\lambda\,2.189$\,\mim\ line (see right
panel in Fig.\,\ref{fig:nv}).

%%%%%%%%%%%%%%%%%%%%%%%%%  FIG 6  %%%%%%%%%%%%%%%%%%%%%%%%%%%%%%%%
%-------------------------------------------------------------
\begin{figure}
\centering
\includegraphics[width=0.9\columnwidth]{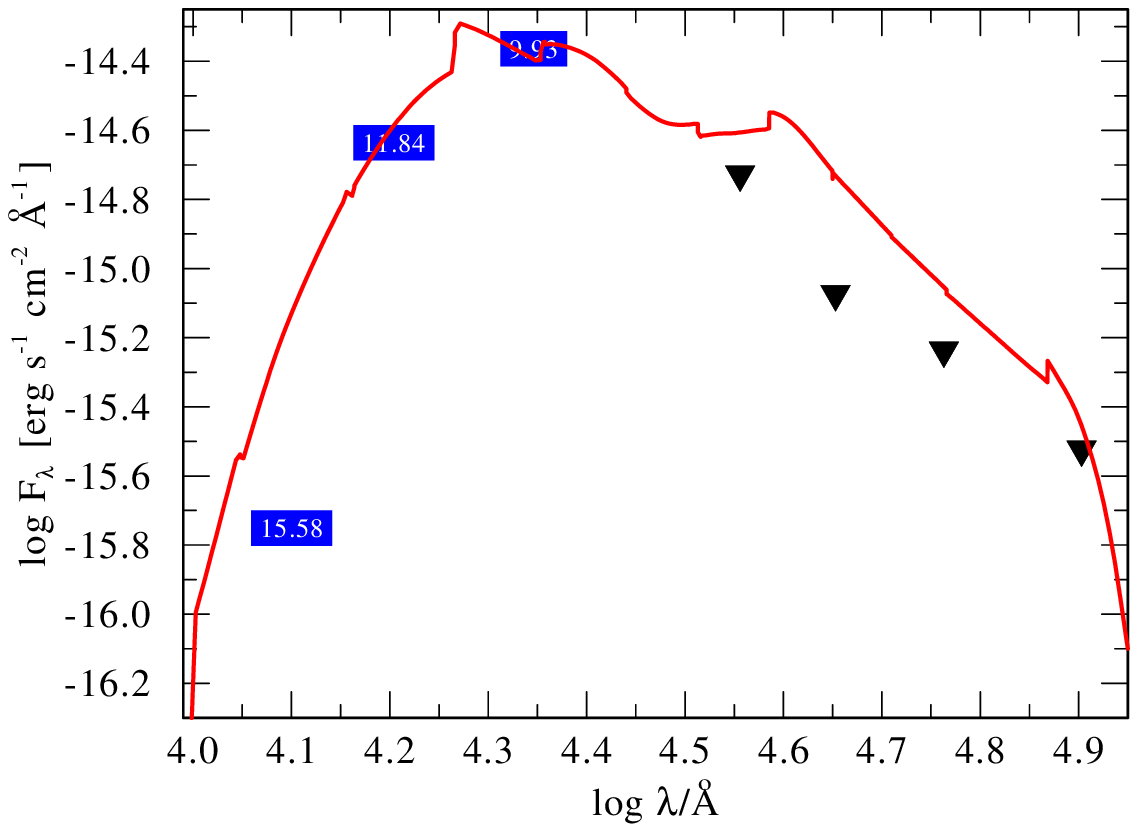}
\caption{Spectral energy distribution for \wrc. The solid line represents the
reddened model flux (see Table\,\ref{tab:par} for the parameters). Boxes give
the observed 2MASS magnitudes (labels). Triangles correspond to the flux
in the four IRAC channels, which we extracted from the archival images
(see Sect.\,\ref{sec:irac} and Table\,\ref{tab:irac}).}
\label{fig:phc}
\end{figure}
%_____________________________________________________________
 
The contribution of the hot companion has been scaled down by a factor
of ten in order to match the weak He\,{\sc ii} and N\,{\sc v} features
in the observation. It is plausible that a hot WNE-type companion would
contribute only little to the composite IR flux. WNE stars are generally
not that luminous, and even for same bolometric luminosity a companion
twice as hot as the primary would be 50\% fainter in the Rayleigh-Jeans
domain of a black-body spectrum. We conclude that \ka\ may be an
Ofpe/WN9+WNE binary system. However, the limited quality of the K-band
data (Fig.\,\ref{fig:k102ka}) does not allow
this question to be settled.

\subsection{Stellar parameters of \wrc}

%%%%%%%%%%%%%%%%%%%%%%%%%  FIG 7  %%%%%%%%%%%%%%%%%%%%%%%%%%%%%%%%
%-------------------------------------------------------------
\begin{figure}
  \centering
  \includegraphics[width=0.9\columnwidth]{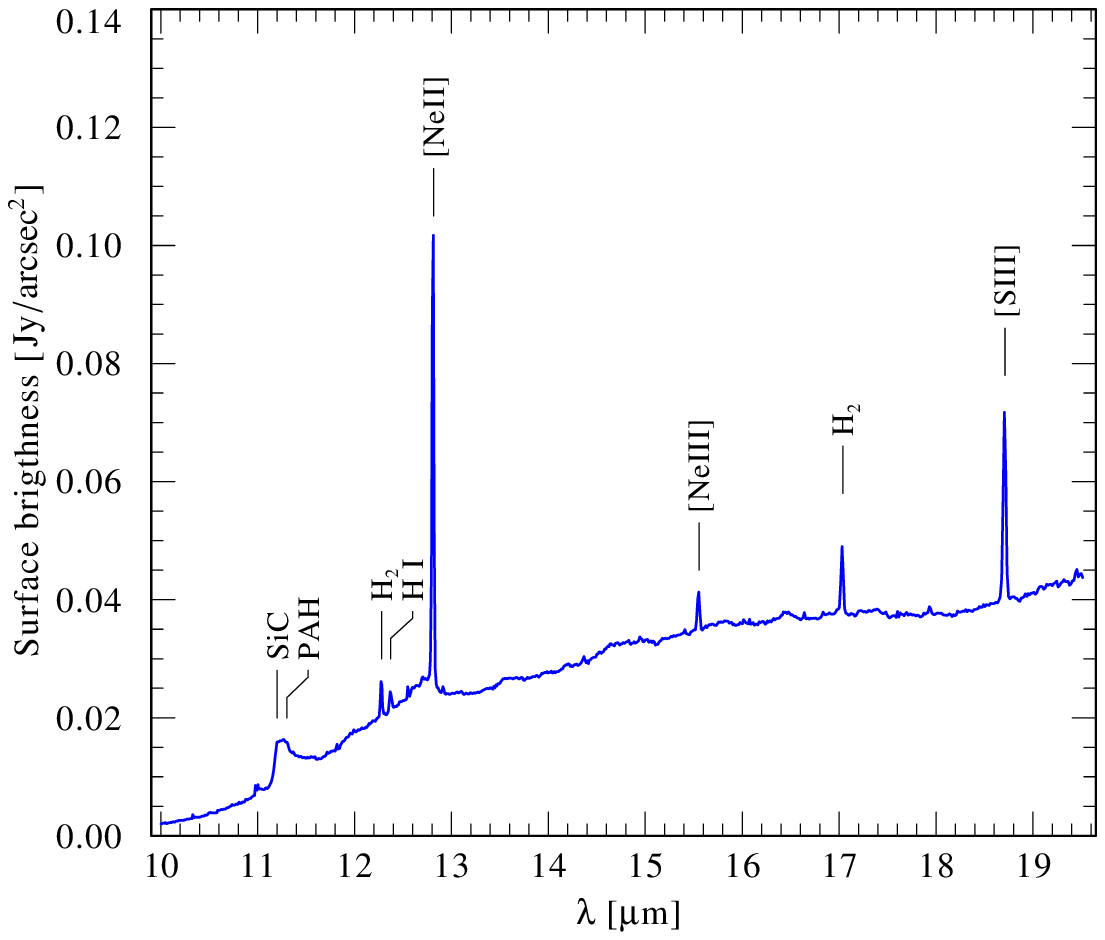}
  \caption{{\em Spitzer} IRS SH spectrum of \ka\ with line
    identifications. Since the emission region is more extended  
than the spectrograph slit, the flux is given in terms of surface 
brightness.}
  \label{fig:ka-irs}
\end{figure}
%
%_____________________________________________________________

%%%%%%%%%%%%%%%%%%%%%%%%%  FIG 8  %%%%%%%%%%%%%%%%%%%%%%%%%%%%%%%%
%-------------------------------------------------------------
\begin{figure}
  \centering
  \includegraphics[width=0.9\columnwidth]{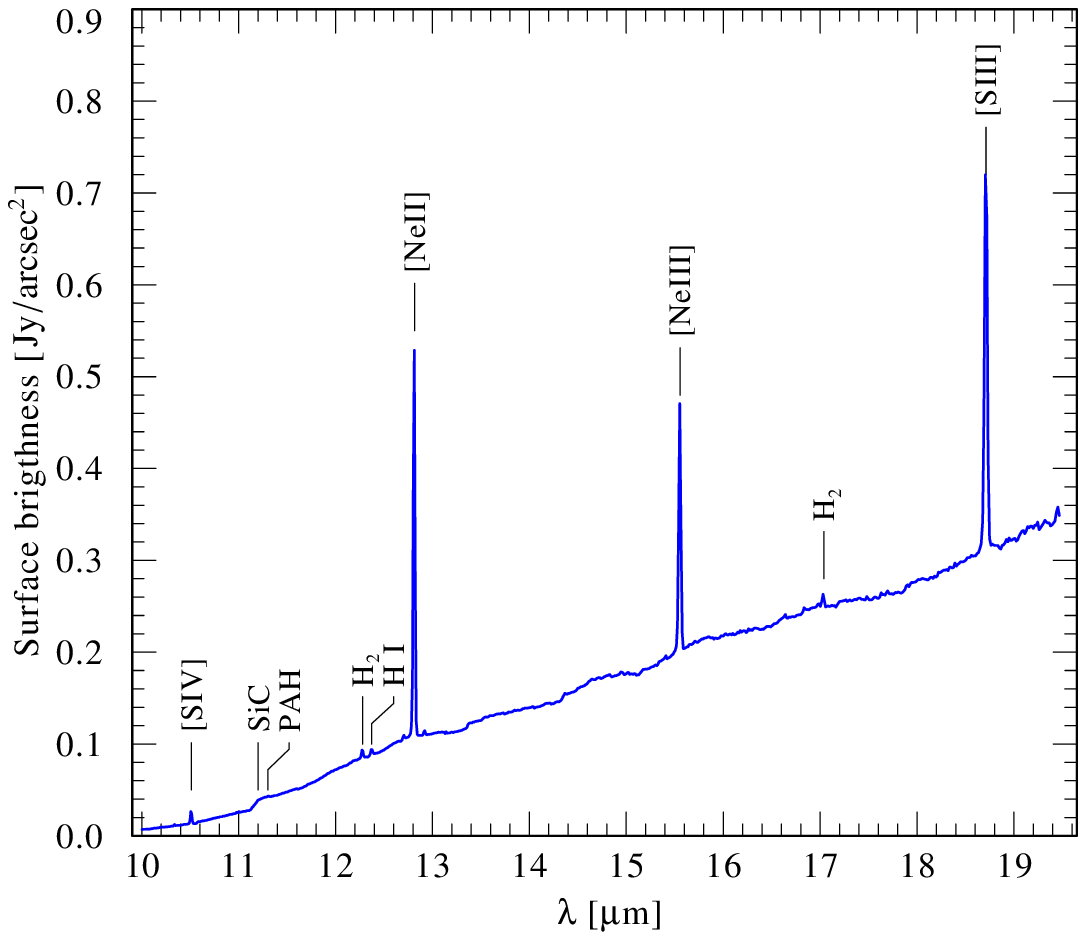}
  \caption{{\em Spitzer} IRS SH spectrum of \wrc\ with line
    identifications. Since the emission region is more extended  
than the spectrograph slit, the flux is given in terms of surface 
brightness.}
  \label{fig:c-irs}
\end{figure}
%  
%_____________________________________________________________

Figer \etal\ (\cite{Fig99}) published a K-band spectrum of \wrc,
unfortunately plotted with an unspecified offset, and {\changed assigned
to it a K-band magnitude of 11.6\,mag. This value is not
consistent with the more recently available 2MASS and the {\em
Spitzer} IRAC point source catalogs. We have to assume that, most
plausibly, the K-band spectrum shown in Figer \etal\ (\cite{Fig99})
in fact belongs to a much brighter star (K-band magnitude 9.93\,mag)
which we identify with \wrc.  This assumption is validated by the
consistent picture which emerges from (i) the photometry marks from
2MASS and {\em Spitzer} IRAC catalogs that are well matched by the
(reddened) spectral energy distribution of a luminous WNL star (see
Figs.\,\ref{fig:phc} and \ref{fig:sedc}), (ii) the location of \wrc\ in the
region of the IRAC color-color diagram populated by
WN stars (Hadfield \etal\ 2007). 

Figer \etal\ (\cite{Fig99}) classified
\wrc\ as a WN6 subtype,} due to the similarity of its
K-band spectrum with WR\,115. Taking this comparison at face value, 
we adopt from the 
analysis of WR\,115 by Hamann \etal\ (\cite{wrh06}) the temperature
$T_\ast\,=\,50\,$kK and terminal wind velocity
$\vinf\,=\,1300\,\rm{km\,s^{-1}}$. To set upper and lower limits to
the temperature, we qualitatively compare the K-band spectrum of \wrc\
with the PoWR models.  We can exclude a $T_\ast$ below 40\,kK, since
cooler models do not show the He\,{\sc ii} emission line at
$\lambda=2.189\,\mim$.  $T_\ast$ above 60\,kK can be excluded because
the He\,{\sc i} singlet line at $\lambda=2.059\,\mim$ is not visible
in hotter models. Despite of the WN-early classification, the relative
strength of the He\,{\sc ii}/Br$\gamma$ blend, compared to un-blended
helium lines, requires a model with some hydrogen ($\sim$20\% by mass
as a very rough estimate). {\changed The WR\,115 comparison star is actually 
hydrogen-free, but apart from the hydrogen lines this has only little 
influence on the stellar spectrum.}  

%%%%%%%%%%%%%%%%%%%%%%%%%  FIG 9  %%%%%%%%%%%%%%%%%%%%%%%%%%%%%%%%
%-------------------------------------------------------------
\begin{figure*}
\centering
\includegraphics[width=13.2cm]{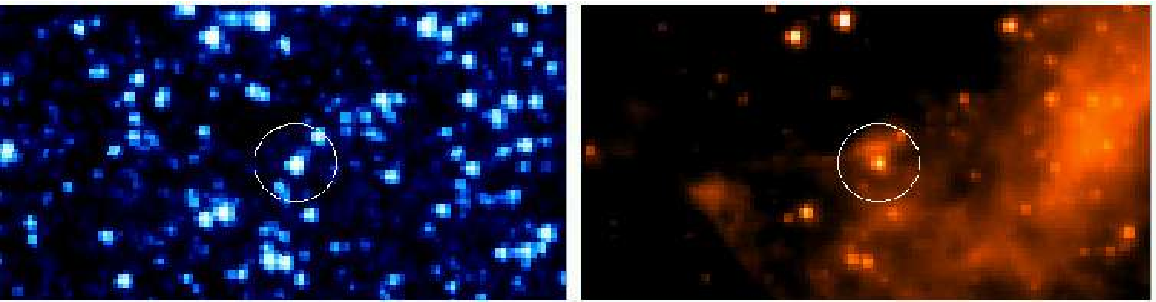}
\caption{{\em Spitzer} IRAC archive image at 3.6\,$\mu\rm{m}$ (left)
  and 8\,$\mu\rm{m}$ (right) of the field around \ka. The image size
  is $\approx 2\farcm 3\times 1\farcm 2$. North is to the top and east
  to the left. The white circles have a radius of 10\,\arcsec\ and are
  centered on the coordinates of \ka.}
\label{fig:irac-ka} 
\end{figure*}
%_____________________________________________________________

%%%%%%%%%%%%%%%%%%%%%%%%%  FIG 10  %%%%%%%%%%%%%%%%%%%%%%%%%%%%%%%%
%-------------------------------------------------------------
\begin{figure*}
\centering
\includegraphics[width=13.2cm]{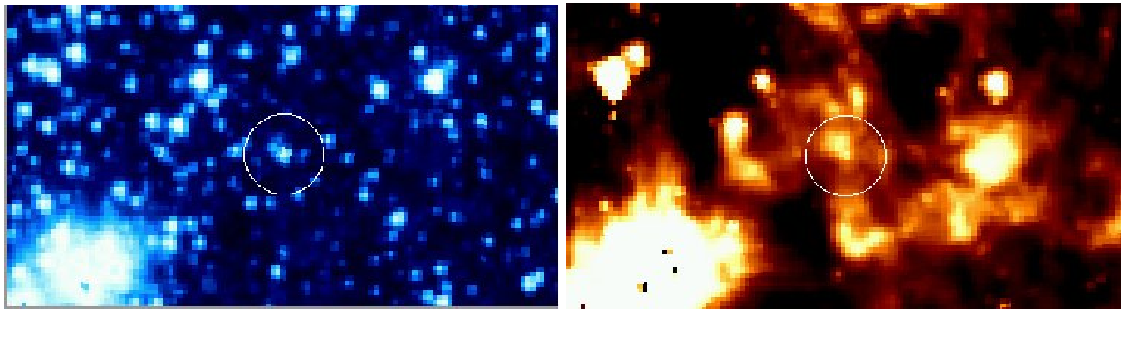}
\caption{
{\em Spitzer} IRAC archive image at 3.6\,$\mu\rm{m}$ (left)
  and 8\,$\mu\rm{m}$ (right) of the field around \wrc. The image size
  is $\approx 2\farcm 3\times 1\farcm 2$. North is to the top and east
  to the left. The white circles have a radius of 10\,\arcsec\ and are
  centered on the coordinates of \wrc.
}
\label{fig:irac-c} 
\end{figure*}
%_____________________________________________________________

{\changed  The luminosity of \wrc\ is derived by scaling the spectral
energy distribution of the model to the photometry marks. The extinction
is simultaneously adjusted, yielding $A_V\,=\,25.7$\,mag. This estimate
is consistent with the extinction in the Sickle nebular region
determined by Cotera \etal\ (\cite{cot00}) and Schultheis \etal\
(\cite{s1999}). The latter work provides an extinction map which} shows
strong spatial variation of $A_V$, typically 1\,mag, in a 1\arcmin\
radius around \wrc. We take this scatter as error estimate for the
adopted $A_V$ value.  In the K-band this corresponds to
$\Delta\,A_K=0.1\,$mag.  Using model grids for WNE and WNL stars, we
inspected the bolometric correction relative to the K-band magnitude for
the range of WN5-WN7 stars analyzed in Hamann \etal\ (\cite{wrh06}),
finding a scatter of about 0.65\,mag which transforms into a luminosity
uncertainty of $\pm 0.26\,$dex. Together with the scatter of
$\Delta\,A_V=1\,$mag from the extinction map, the total uncertainty in
luminosity amounts to $\pm 0.3\,$dex. {\changed The derived luminosity
of \wrc, $\log(L)\,[L_\odot ]\approx 6.3$, is typical for WNL-type stars
(see Hamann \etal\ \cite{wrh06}).}

Figure\,\ref{fig:phc} shows that the correspondingly reddened stellar
model flux does not perfectly fit to the 2MASS and IRAC photometric
observations.  However, it should be noted that \wrc\ resides in a
crowded sky region which makes photometry measurements difficult. For
instance, the quality flags in the 2MASS All Sky Catalog of Point
Sources indicate difficulties in the photometry determination. Since
we fit our synthetic SED to these photometric observations, this
induces some uncertainty to the derived luminosity. Some further error
margins arise from the poorly constrained model parameters. The
adopted stellar parameters of \wrc\ are listed in
Table\,\ref{tab:par}.

%%%%%%%%%%%%%%%%%%%%%%%%%%%%%%%%%%%%%%%%%%%%%%%%%%%%%%%%%%%%%%%%%%%%%%%%%%%%
%%%%%%%%%%%%%%%%%%%%%%%%%%%%%%%%%%%%%%%%%%%%%%%%%%%%%%%%%%%%%%%%%%%%%%%%%%%%
%--------------------  Section 2 Observations -----------------------------%
%%%%%%%%%%%%%%%%%%%%%%%%%%%%%%%%%%%%%%%%%%%%%%%%%%%%%%%%%%%%%%%%%%%%%%%%%%%%
%%%%%%%%%%%%%%%%%%%%%%%%%%%%%%%%%%%%%%%%%%%%%%%%%%%%%%%%%%%%%%%%%%%%%%%%%%%%

\section{Mid-IR observations and data reduction}

We obtained IR spectra of \ka\ and \wrc\ with the IRS spectrograph of
the {\em Spitzer} Space Telescope (the observation log is shown in
Table\,\ref{table:obs}).  In addition, our program stars were observed
by various imaging mid-IR instruments ({\em Spitzer} IRAC, {\em
Spitzer} MIPS, and MSX) that performed surveys of the GC. These
observations are briefly described below.

\subsection{{\em Spitzer} IRS spectra}

The {\it Spitzer} Infrared Spectrograph (IRS) Short-High (SH) module
covers the wavelength range 9.9\,-\,19.6\,\mim\ and provides a
moderate spectral resolving power of $R\!\approx\!600$ with a slit
aperture of $4\farcs 7 \times 11\farcs 2 $ (Houck
\etal\ \cite{houck04}). 

%-------------------------------------------------------------
\begin{table}
\caption{{\em Spitzer} IRS observations of \ka\ and \wrc.}
\label{table:obs}
\centering
\begin{tabular}{ccc}
\hline
\hline
\rule[0mm]{0mm}{3.25mm}                  & WR 102ka     & WR 102c         \\
\hline                                                                
%Observing Mode & IRS SH staring & IRS S               H staring \\    
\rule[0mm]{0mm}{3.25mm} RA J2000   & 17$^{\rm h}$\,46$^{\rm m}$\,18\fs 12  & 
                                     17$^{\rm h}$\,46$^{\rm m}$\,10\fs 91   \\
DEC J2000                         & $-29\degr\,01\arcmin\,36\farcs 6$  
                                  & $-28\degr\,49\arcmin\,07\farcs 4$     \\
Integration time [s]                     &  4145        & 2926            \\
%Ramp duration [s]& 120 & 120  \\                                             
Number of cycles                         & 12           & 17              \\
AOR ID                                   & 10878720     & 10878976        \\
\hline 
\rule[0mm]{0mm}{3.25mm} Program ID       & \multicolumn{2}{c}{3397}       \\
\rule[0mm]{0mm}{3.25mm} Observation date & \multicolumn{2}{c}{20/04/2005} \\
\hline
\end{tabular}
\end{table}
%-------------------------------------------------------------

Already the first inspection of the pipe-line extracted spectra of
both program stars revealed a strong mid-IR flux (far above the flux
expected from the synthetic stellar SEDs) increasing towards longer
wavelength.  The intensity along the slit does not vary between the
two ``nodding'' positions (in each exposure cycle, the target is
displaced towards one or the other end of the slit in turn).
Therefore we conclude that in both cases the mid-IR emission is not
due to a stellar point source, but emerges from an extended area
larger than the spectrograph slit.  Hence we extracted the IRS spectra
(using the {\it SPICE} ver.\ 1.3 software) under the assumption of an
extended emission region with uniform surface brightness.

After the data reduction with {\it SPICE}, the individual echelle
orders of the IRS spectra do not match at their wavelength
overlap. For cosmetic reasons we slightly tilted each spectral order
until they fit perfectly.  The resulting spectra are shown in
Figs.\,\ref{fig:ka-irs} and
\ref{fig:c-irs}. The spectra display prominent forbidden 
emission lines.  Relatively weaker lines of H$_2$, H\,{\sc i},
He\,{\sc ii} as well as SiC and/or PAH features are also present. The
equivalent widths and the line strengths of the individual emission
lines are compiled in Table\,\ref{tab:irs} for both objects.

The immediate question is whether the IRS spectra are merely dominated
by the GC background or do we indeed measure emission from
circumstellar nebulae physically associated with the program stars.
To answer this key question we retrieve and examine all available
mid-IR images of \ka\ and \wrc.
 
%---------------------------------------------------------------
\begin{table*}
  \caption{Equivalent widths, 
    intensities, and intensity ratios of the prominent emission lines in   
    the IRS spectra of \ka\ and \wrc}
\label{tab:irs}
\centering
\begin{tabular}{l
D{.}{.}{-1}
D{.}{.}{-1}
D{.}{.}{-1}
D{.}{.}{-1}
D{.}{.}{-1}
}
\hline
\hline
& & 
\multicolumn{2}{c}{\rule[0.5ex]{1cm}{0.1mm} \ka\ \rule[0.5ex]{1cm}{0.1mm} } & 
\multicolumn{2}{c}{\rule[0.5ex]{1cm}{0.1mm} \wrc \rule[0.5ex]{1cm}{0.1mm}}\\

\rule[0mm]{0mm}{3.25mm}   
& \multicolumn{1}{c}{Wavelength} 
& \multicolumn{1}{c}{$-\rm{W_{\lambda}}$}  
& \multicolumn{1}{c}{$I_{\rm line}$} 
& \multicolumn{1}{c}{$-\rm{W_{\lambda}}$}  
& \multicolumn{1}{c}{$I_{\rm line}$}   \\

& \multicolumn{1}{c}{[\mim]}     
& \multicolumn{1}{c}{[\mim]}
& \multicolumn{1}{c}{[$10^{-4}$\,erg\,s$^{-1}$\,cm$^{-2}$\,sr$^{-1}$]}
& \multicolumn{1}{c}{[\mim]}
& \multicolumn{1}{c}{[$10^{-4}$\,erg\,s$^{-1}$\,cm$^{-2}$\,sr$^{-1}$]} \\

\hline
\rule[0mm]{0mm}{3.25mm}$[$S\,{\sc iv}$]$   & 10.51  & <\,0.002 & <0.0085 & 0.019 & 2.94 \\ 
H$_2$\,[J=4-2]\,S(2)  & 12.28 & 0.0062     & 1.06   & 0.0021   & 1.49 \\
H\,{\sc i}\,7\,--\,6            & 12.37 & 0.0036     & 0.64   & 0.0016   & 1.19 \\ 
$[$Ne\,{\sc ii}$]$    & 12.81  & 0.073      & 14.7  & 0.084    & 73.2 \\
$[$Ne\,{\sc iii}$]$   & 15.55 & 0.0055     & 1.02   & 0.036   & 38.0 \\
H$_2$\,[J=3-1]\,S(1)  & 17.04 & 0.0081     & 1.32   & 0.0013   & 1.40 \\
$[$S\,{\sc iii}]$$    & 18.7  & 0.028      & 4.04   & 0.047    & 54.0 \\
\hline %----------------------------------------------------------------
\rule[0mm]{0mm}{3.25mm}$[$S\,{\sc iv}$]\,/\,[$S\,{\sc iii}$]$ &  
& \multicolumn{2}{D{.}{.}{-1}}{< 0.002} 
& \multicolumn{2}{D{.}{.}{-1}}{ 0.054 }\\

$[$Ne\,{\sc iii}$]\,/\,[$Ne\,{\sc ii}$]$ &  
& \multicolumn{2}{D{.}{.}{-1}}{0.07} 
& \multicolumn{2}{D{.}{.}{-1}}{0.52}\\

$[$Ne\,{\sc iii}$]\,/\,[$S\,{\sc iii}$]$ &  
& \multicolumn{2}{D{.}{.}{-1}}{0.25} 
& \multicolumn{2}{D{.}{.}{-1}}{0.76}\\

H$_2$-S(1)\,/\, H$_2$-S(2) &  
& \multicolumn{2}{D{.}{.}{-1}}{1.25} 
& \multicolumn{2}{D{.}{.}{-1}}{0.94}\\
\hline %--------------------------------------------------------
\end{tabular}
\end{table*}

\subsection{IRAC and MSX images and photometry of \ka\ and \wrc}
\label{sec:irac}

{\em Spitzer} IRAC observations of the central part of the Galaxy were
presented by Stolovy \etal\ (\cite{sto06}). IRAC has high angular
resolution (pixel size  $\sim 1\farcs 2\,$) and sensitivity (Fazio
\etal\ \cite{irac}). It provides simultaneous 
$5\farcm 2\times 5\farcm 2$ images in four channels (cf.\ Table\,\ref{tab:irac}).

We have retrieved and analyzed the archival IRAC images of the fields 
containing \ka\ and \wrc.  For illustration, Figs.\,\ref{fig:irac-ka}
and \ref{fig:irac-c} show these images in the first and the forth IRAC
channel. Both \wrc\ and \ka\ are detected at all four wavelengths. To
extract photometric fluxes the current version of the {\em mopex}
software was used (see Table\,\ref{tab:irac}). For the IRAC magnitudes
and colors we adopted the calibration by Reach \etal\ (\cite{iracm}).
The IRAC color indices of \ka\ and \wrc\ agree well with those of
other galactic WR stars as presented by Hadfield \etal\ (\cite{had07})
on basis of the IRAC GLIMPSE survey.

\begin{figure}
\centering
\includegraphics[width=0.9\columnwidth]{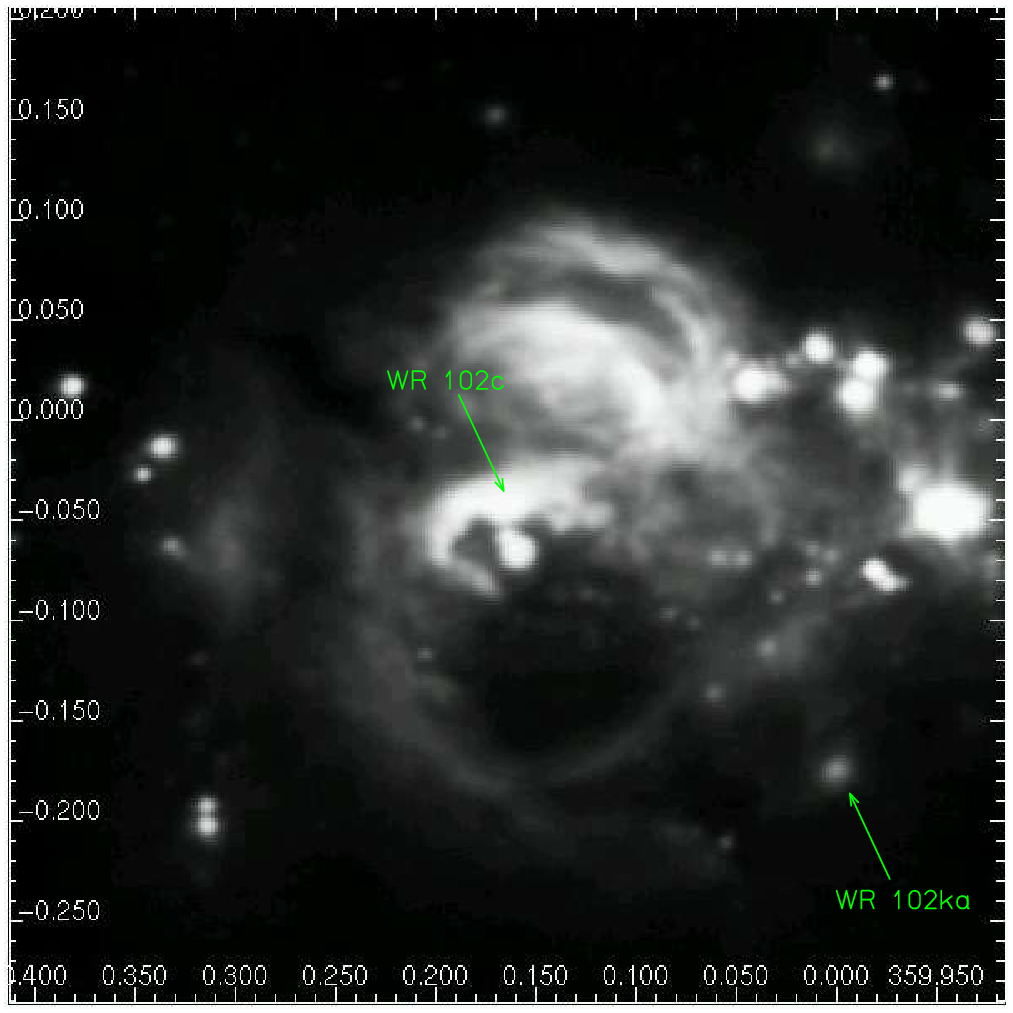}
\caption{MSX band E image (histogram  equalization  scale). 
Positions of nebulae around \ka\ and \wrc\ are indicated by the arrows.}
\label{fig:msx}
\end{figure}

Images of the \ka\ and \wrc\ fields at  wavelengths longer than those
covered by IRAC were obtained with the Midcourse Space Experiment
(MSX). The intrinsic angular resolution of this instrument -- limited
by its pixel size -- is only 20\,\arcsec, which corresponds to 0.8\,pc
linear extent at the distance of the GC (Price \etal\ \cite{msx01}).
Figure\,\ref{fig:msx} displays the archival MSX E-band image of the
GC region. Both our objects show up as bright structures
far above the general background, albeit in different morphological
context.

\begin{figure*}
   \centering
   \includegraphics[width=13.2cm]{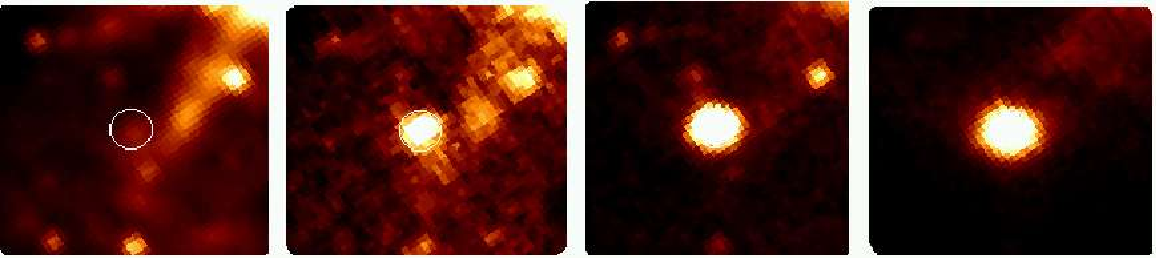}
   \caption{Archival MSX images of the sky field around \ka, from left to
     right in the  A-band (8.28\,\mim), C-band (12.13\,\mim), 
     D-band (14.65\,\mim),
     and E-band (21.34\,\mim). Each image has a size of $\approx
     4$\arcmin$\times 4$\arcmin. North is to the top and 
       east to the left.  The circles have a radius of 20\,\arcsec\ and are
     centered on the coordinates of \ka.}
         \label{fig:msx-ka}
   \end{figure*}

\ka\ coincides with an isolated, unresolved source visible in the
wavelength bands C, D and E (see Fig.\,\ref{fig:msx-ka}),  as already
noticed by Clark \etal\ (\cite{lbv2005}). The fluxes from the MSX point
source catalog are included in Table\,\ref{tab:msx}.

%%%%%%%%%%%%%%%%%%%%%%%%%  FIG 11  %%%%%%%%%%%%%%%%%%%%%%%%%%%%%%%%
%-------------------------------------------------------------
%
%_____________________________________________________________

\begin{figure*}
\centering
\includegraphics[width=13.2cm]{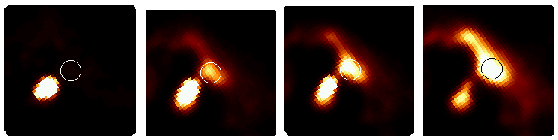}
\caption{Archival MSX images of the sky field around \wrc,
  from left to right in the A-band (8.28\,\mim), C-band (12.13\,\mim), D-band
  (14.65\,\mim), and E-band (21.34\,\mim). Each image has a size of
  $\approx 4$\arcmin$\times 4$\arcmin. North is to the top and
  east to the left. The circles have a radius  of 20\,\arcsec\ and are
  centered on the coordinates of \wrc. The Quintuplet cluster can be seen 
  in the lower left part of the images.}  
\label{fig:msx-c}
\end{figure*}

\wrc\ resides in an IR-bright, extended H\,{\sc ii} region, the Sickle
nebula.  Figure\,\ref{fig:msx-c} shows its MSX images. The unresolved
Quintuplet cluster is the dominant source in the A-band (8.28\mim).
However, towards longer wavelengths the emission from the Sickle nebula
embedding \wrc\ is strongly increasing. Remarkably, in the E-band
(18.2-25.1\mim) the region centered on \wrc\ is significantly brighter
than the whole Quintuplet cluster. The Sickle nebula appears as an
extended complex even with the poor resolution of MSX. The MSX point
source catalog lists a couple of sources that coincide with \wrc\
within the resolution. The closest entry, G000.1668-00.0434, is offset
from the 2MASS coordinates of \wrc\ by 7\arcsec. Somewhat arbitrarily
we identify this source with \wrc\ and adopt the corresponding fluxes
(cf.\ Table\,\ref{tab:msx}). 

The MSX bands cover the wavelength range of our {\em Spitzer} IRS 
spectra. Because both program stars are detected as MSX point sources 
in all bands we conclude that the IRS spectra probe emission 
originating from nebulae physically associated with \ka\ and \wrc.  

For MSX with its 20\arcsec$\times$20\arcsec\ pixels these  objects 
remain unresolved point sources. However, we had concluded   that the
emission sources are more extended than the {\em Spitzer} IRS slit of
$4\farcs 7\times 11\farcs 2$.  When multiplying the measured surface
flux with the slit area, we obtain the flux covered by the slit and
plot it in Figs.\,\ref{fig:sed}  and \ref{fig:sedc}. The MSX point
source fluxes lie higher by a factor 5 to 7 for both program stars.
This is consistent with the above conclusion that both sources are
more extended than the area covered by the IRS slit.

%---------------------------------------------------------------
\begin{table}
\caption{IRAC fluxes, magnitudes, and IR colors of \ka\ and \wrc}
\label{tab:irac}
\centering
\begin{tabular}{cccc}
\hline
\hline
\rule[0mm]{0mm}{3.25mm} IRAC  & Wavelength  &  WR\,102ka  &  WR\,102c               \\
                     channel  &    & Flux~/~magnitude      & Flux~/~magnitude       \\
                  & [\mim]      & [Jy]~/~[mag$^{\rm a}$]  & [Jy]~/~[mag$^{\rm a}$]  \\
\hline                       
\rule[0mm]{0mm}{3.25mm}  1    & 3.6         & 0.211~/~7.81          & 0.081~/~8.8    \\
                         2    & 4.5         & 0.214~/~7.31          & 0.058~/~8.7    \\
                         3    & 5.8         & 0.174~/~7.05          & 0.065~/~8.12   \\
                         4    & 8.0         & 0.112~/~6.89          & 0.064~/~7.50   \\
\hline            
\multicolumn{2}{l}{[3.6] - [8.0]}           & 0.92                  & 1.35          \\
\multicolumn{2}{l}{[3.6] - [4.5]}           & 0.50                  & 0.12          \\
\multicolumn{2}{l}{[5.8] - [8.0]}           & 0.16                  & 0.62         \\
\hline
\multicolumn{4}{l}
{\rule[0mm]{0mm}{3.25mm}a: IRAC magnitude system from 
Reach \etal\ (\cite{iracm})}\\
\end{tabular}
\end{table}

%---------------------------------------------------------------

%%%%%%%%%%%%%%%%%%%%%%%%%  FIG 12  %%%%%%%%%%%%%%%%%%%%%%%%%%%%%%%%
%-------------------------------------------------------------
%
%_____________________________________________________________

%%%%%%%%%%%%%%%%%%%%%%%%%  FIG 13  %%%%%%%%%%%%%%%%%%%%%%%%%%%%%%%%
%-------------------------------------------------------------
%
%_____________________________________________________________

%---------------------------------------------------------
\begin{table}
\caption{MSX photometry of \ka\ and \wrc\ from the MSXC6 Catalog}
\label{tab:msx}
\centering
\begin{tabular}{c
D{.}{.}{-1}
D{.}{.}{-1}
D{.}{.}{-1}
}
\hline
\hline
\rule[0mm]{0mm}{3.25mm}MSX band & \multicolumn{1}{c}{Wavelength} & 
\multicolumn{1}{c}{Flux (WR\,102ka)}&  \multicolumn{1}{c}{Flux (WR\,102c)}    \\
 & \multicolumn{1}{c}{[\mim]}    & \multicolumn{1}{c}{[Jy]}            
&  \multicolumn{1}{c}{[Jy]}              \\
\hline                                                                    
\rule[0mm]{0mm}{3.25mm}A         & 8.28      & 0.59            & 1.29              \\
C                                & 12.13     & 5.39            & 23.35             \\
D                                & 14.65     & 12.38           & 61.98             \\
E                                & 21.34     & 30.77           & 207.50            \\
\hline 
\end{tabular}
\end{table}
%------------------------------------------------------------

%%%%%%%%%%%%%%%%%%%%%%%%%%%%%%%%%%%%%%%%%%%%%%%%%%%%%%%%%%%%%%%%%%%%%%%%%%%%%
%--------------------  Section 4 Analysis if IR spectra --------------------
%%%%%%%%%%%%%%%%%%%%%%%%%%%%%%%%%%%%%%%%%%%%%%%%%%%%%%%%%%%%%%%%%%%%%%%%%%%%%

\section{Analysis of mid-IR spectra and images \label{sec:irana}}

\subsection{Temperature and mass of the dust 
around \ka\ and \wrc\ \label{seq:dust}}

As illustrated in Figs.\,\ref{fig:msx-ka},  \ref{fig:msx-c} and
Table\,\ref{tab:msx}, the flux is increasing towards longer wavelengths
and is strongest in the E band for both objects.  The emission in this
band is mostly due to dust grains heated by starlight  (Cohen \& Green
\cite{cg01}). 

In order  to determine the temperature, the mass, and the composition
of the circumstellar dust around \ka\ and \wrc, we use the publicly
available code {\sc dusty} (Ivezi\'c \& Elitzur \cite{E1997})  for
modeling the continuum emission. This code treats the continuum
radiative transfer in dust for a spherical circumstellar nebula,
irradiated by a central star with a given emergent flux. For the optical
properties of the dust, the theoretical grain model from Draine\,\etal\
(\cite{dr84}) is implemented. By suitable transformation to scale-free
quantities, {\sc dusty} requires the following free parameters to be
specified: 

\noindent
{\em i) The radiation field of the central star.} This parameter 
is determined by model stellar spectra obtained from our spectral 
analyses (Sect.\,\ref{sect:analysis}).\\
\noindent
{\em ii) $T_1$, the temperature at the inner boundary. }
This parameter mainly influences the wavelength of flux maximum. {\sc
dusty} assumes that the dust temperature is in radiative equilibrium
with the radiation field. Hence, $T_1$ implicitly also fixes the inner
radius of the dust shell, $r_1$.\\
\noindent
{\em iii) $Y=r_2/r_1$, the radial extent of the dust shell.} This
parameter gives the outer boundary $r_2$ in units of the inner
boundary $r_1$. $Y$ influences the temperature stratification. In the
optically thin case (which is relevant for our objects), a
thin shell is nearly isothermal.\\
\noindent
{\em iv) The radial density profile of the dust shell.}
For simplicity, we assume $\rho(r) \propto r^{-2}$ for stationary 
expansion at constant velocity.\\
\noindent
{\em v) $\tau_V$, the radial optical depth of the 
dust shell at 5500\,\AA.}
This parameter governs the brightness of the dust emission. In the 
optically thin regime, the fraction of stellar radiation which is 
converted by the dust into IR emission depends roughly linearly on 
$\tau_V$.\\
\noindent
{\em vi) The grain size distribution.}
We adopt a usual power-law distribution, $n_{\rm dust}(a) \propto
a^{-q}$, for the grain size $a$, parameterized by the exponent $q$ and
limits for the smallest and largest dust grain diameter, $a_{\rm min}$
and $a_{\rm max}$. For the exponent we take $q=3.5$ after Mathis \etal\ 
(\cite{m1977}). The cut-off values influence the temperature and optical
properties of the dust.\\
\noindent
{\em vii) The composition of dust.} We assume a chemical composition as
  usually adopted for the standard ISM 
(Draine\,\cite{dr04}), i.e.\ 47\% of graphite and 53\% of silicate grains.

%%%%%%%%%%%%%%%%%%%%%%%%%  FIG 14 %%%%%%%%%%%%%%%%%%%%%%%%%%%%%%%%
%-------------------------------------------------------------
\begin{figure*}
\centering
\includegraphics[width=0.8\textwidth]{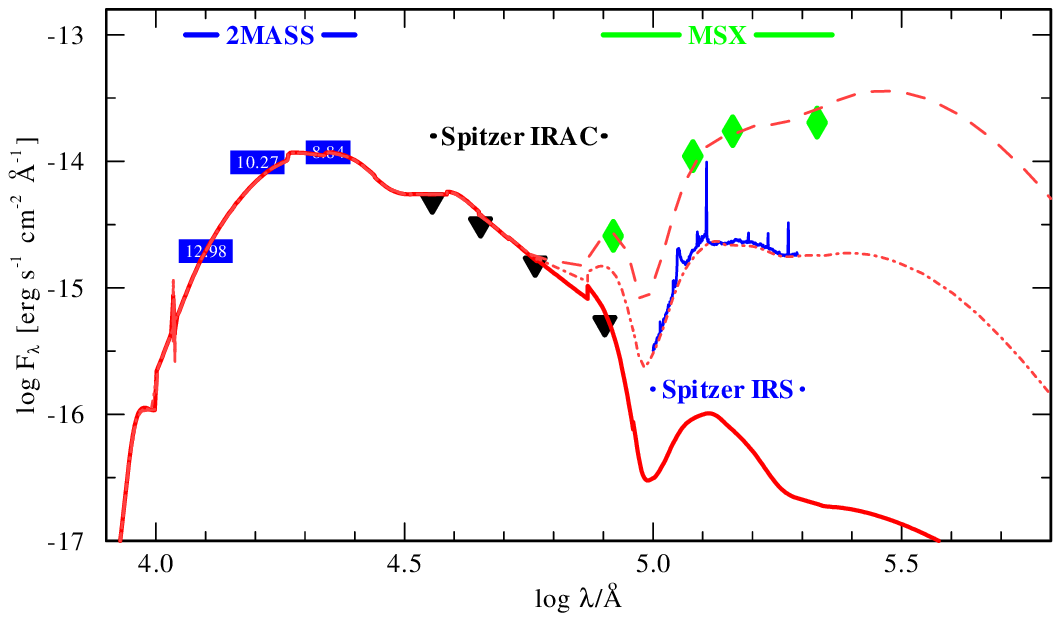}
\caption{Spectrum of \ka. Boxes, triangles and diamonds correspond to
photometric measurements with 2MASS, {\em Spitzer} IRAC, and MSX, respectively 
(cf.\ Tables\,\ref{tab:irac} and \ref{tab:msx}).  
The thin blue line is the spectrum observed with {\em Spitzer} IRS. 
It contains only the flux from those parts of the extended object which 
have been covered by the spectrograph slit. 
The thick red line gives the flux of our stellar model. The model of the 
circumstellar shell adds dust emission. The dash-dotted line represents 
emission from the inner part of the circumstellar nebula that would be covered 
by the {\em Spitzer} IRS slit, while the long-dashed line contains the 
simulated emission from the whole object. }  
\label{fig:sed}
\end{figure*}
%
%_____________________________________________________________

A series of models with various choices of the parameters $T_1$, $Y$,
$\tau_{\rm V}$ and grain size ($a_{\rm min}$, $a_{\rm max}$) were
computed and fitted to the observed IR SED of our program stars.

The cut-off values of the grain size distribution, $a_{\rm min}$ and
$a_{\rm max}$, have strong influence on the dust spectral energy
distribution.  Lowering $a_{\rm min}$ increases the amount of flux
blueward 8.3\,\mim.  Very small grains (VSG) with diameter
$\lsim\,0.001$\,\mim, increase the portion of scattered light and heat
up more rapidly by the absorption of one energetic photon compared to
larger grains.  The {\sc dusty} code, however, does not account
for the detailed physics of the VSG.  The upper cut-off value $a_{\rm
max}$ is difficult to constrain since the large grains contribute
mostly to longer wavelengths where no data are available.

%--------------------------------------------------------------------------
\begin{table}
\caption{Parameters of the dust  models for  \ka\ and  \wrc. The ``inner part''
model is designed to reproduce the SED as observed through the
aperture of the {\em Spitzer} IRS spectrograph, while the emission
from the ``outer part'' model matches the SED constrained by
the MSX photometric measurements  (see text for details).}

\label{tab:dustparameters}
\centering
\begin{tabular}{l|cc|cc}
\hline
\hline
\rule[0mm]{0mm}{3.25mm}  & \multicolumn{2}{c|}{\ka}  & \multicolumn{2}{c}{\wrc}\\
\hline
\rule[0mm]{0mm}{3.25mm}Input parameter  &  &  & & \\
\hline
$a_{\rm{min}}\,[\mim ]$ & \multicolumn{2}{c|}{0.001}     & \multicolumn{2}{c}{0.005} \\
$a_{\rm{max}}\,[\mim ]$ & \multicolumn{2}{c|}{7.50}      & \multicolumn{2}{c}{10.0}  \\
\hline
\rule[0mm]{0mm}{3.25mm} Input parameter &\multicolumn{1}{c}{inner part} & 
\multicolumn{1}{c|}{outer part} & \multicolumn{1}{c}{inner part} & 
\multicolumn{1}{c}{outer part}\\
\hline
\rule[0mm]{0mm}{3.25mm} $T_{1}\,[\rm{K}]$   & 200   & 150      &  175           & 130\\
$Y$                      &  2                            & 3                            &
 2              & 3\\
$\tau_{\rm{V}}$         &$1.2\cdot 10^{-3}$ & $1.7\cdot 10^{-2}$    &$8.1\cdot 10^{-3}$    
& $1.5\cdot 10^{-1}$\\
\hline
\rule[0mm]{0mm}{3.25mm} Inferred parameter &\multicolumn{1}{c}{inner part} & 
\multicolumn{1}{c|}{outer part} & \multicolumn{1}{c}{inner part} & 
\multicolumn{1}{c}{outer part}\\
\hline
\rule[0mm]{0mm}{3.25mm} $r_1\,[\rm{pc}]$ & 0.06      & 0.12        &0.07       & 0.15\\
\rule[0mm]{0mm}{3.25mm} $r_2\,[\rm{pc}]$ & 0.12      & 0.36        &0.15  & 0.45\\
$\tau_{20\,\mim}$       &$1.6\cdot 10^{-4}$ & $2.2\cdot 10^{-3}$ &$1.2\cdot 10^{-3}$ &
$1.3\cdot 10^{-2}$\\ 
$M_{\rm{dust}}\,[\Msun ]$  &$6\cdot 10^{-5}$ & $5\cdot 10^{-3}$ &  $7\cdot 10^{-4}$ & 
$8\cdot 10^{-2}$ \\
\hline
\end{tabular}
\end{table}
%------------------------------------------------------------------

As it was mentioned earlier, the inner parts of the dusty nebulae
around our program stars are covered by the {\em Spitzer} IRS slit
(0.18\,pc\,$\times$\,0.43\,pc at the distance of the GC), while the
MSX pixels correspond to a larger area (0.8\,pc\,$\times$\,0.8\,pc).
The {\sc dusty} code provides only the total flux and cannot be used
to compute angle-dependent intensities. Therefore we approximate an
extended nebula by stacking together two {\sc dusty} models: an
``inner part'' that roughly fits into the {\em Spitzer} slit, and an
``outer part'' that adds to the total flux observed with MSX.  While
the temperature $T_1$ in the ``inner part'' model is a parameter, for
the ``outer part'' model $T_1$ is fixed by the temperature profile in
the dust shell, such that there is continuous temperature distribution
across both the ``inner'' and the ``outer'' part.

\begin{figure*}
\centering
\includegraphics[width=0.8\textwidth]{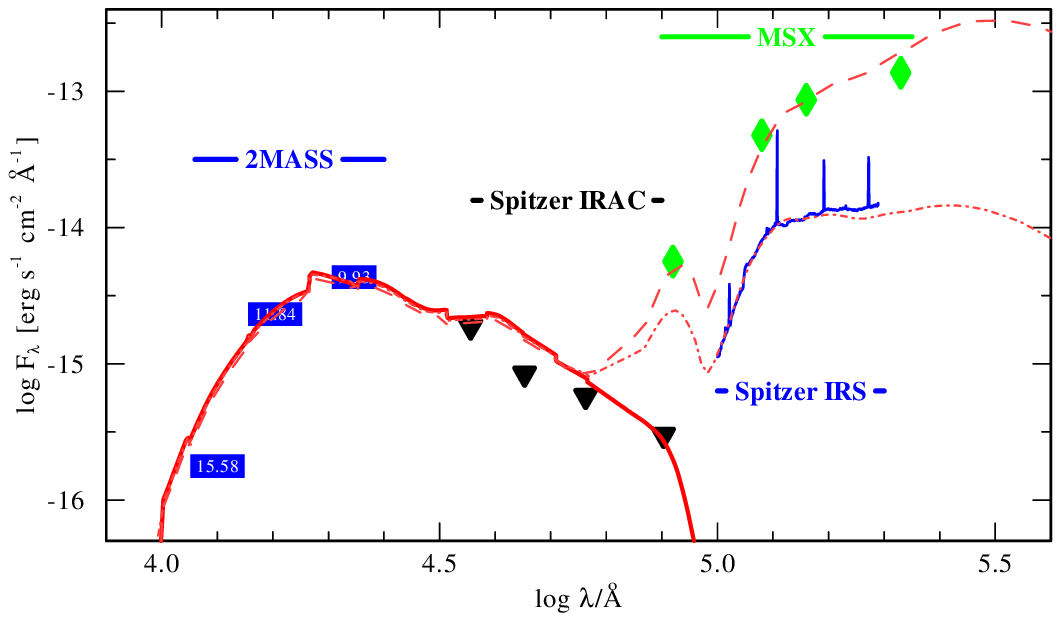}
\caption{Synthetic stellar spectrum (WNE subtype) of \wrc\ (thick red line).
  Boxes, triangles and diamonds correspond to photometric measurements with
  2MASS, {\em Spitzer} IRAC, and MSX, respectively (cf.\
  Tables\,\ref{tab:irac} and \ref{tab:msx}).  The thin blue line is the
  spectrum observed with {\em Spitzer} IRS.  It contains only the flux from
  those parts of the extended object which have been covered by the
  spectrograph slit. The model of the circumstellar shell adds dust emission.
  The dash-dotted line represents emission from the inner part of the
  circumstellar nebula that would be covered by the {\em Spitzer} IRS slit,
  while the long-dashed line contains the simulated emission from the whole
  object.}
\label{fig:sedc}
\end{figure*}

Figures\,\ref{fig:sed} and \ref{fig:sedc} display our best fit to the
measurements. The input model parameters and  inferred quantities are
compiled  in Table\,\ref{tab:dustparameters}. The ``inner part'' model
fits well to the continuum emission  in the IRS aperture, while the
MSX photometry is reproduced by the co-added flux from both the
``inner'' and the ``outer'' part.

Our analysis reveals dust remarkably close to the WN stars. In case of
\ka\ the dust is found as close as 1000\,$R_\ast$ from the stellar
surface! In case of the hotter \wrc\, the inner boundary of the dust
envelope is at 5000\,$R_\ast$.  To our knowledge, this is the first
detection of dust in such close proximity to a WN-type star.

To infer the mass of dust we first compute the density using the
information on the dust opacity.  According to Li (\cite{Li05}), the
mass absorption coefficient of the Draine
\etal\ (\cite{dr84}) grain model is given by 
$\chi_{\rm{abs}}\approx 4.6\cdot 10^5 (\lambda/\mim)^{-2}\,\rm{cm^2\,g^{-1}}$. 
For $\lambda=20$\,\mim\ this yields 
$\chi_{\rm{abs}}\approx 1150\,\rm{cm^2\,g^{-1}}$.  
It should be remembered that the absolute value of the mass absorption
coefficient depends strongly on the underlying grain model and can
differ in extreme cases  by an order of magnitude.

We denote the dust opacity by
$\kappa(r)$$=$$\rho(r)\chi_{\rm{abs}}$. The inverse square dilution of
the density yields  $\rho(r)$$=$$\rho_1\,(r_1/r)^2$. The radial
optical depth is $\tau_{20\,\mim}$$=$$
\int^{r_2}_{r_1}\kappa(r)\,{\rm d}r$$=$$\int^{r_2}_{r_1} \rho
(r)\,\chi_{\rm{abs}}{\rm d}r$. Integrating and rearranging for the density
results in 
\begin{equation} 
\rho_1 = \frac{\tau_{20\,\mim}}{r_1\,\chi_{\rm{abs}}(1-Y^{-1})}.
\end{equation}
The dust mass is given by the integral over the volume, i.e.\ $M_{\rm
dust}$$=$$4\pi\,\rho_1\,r_1^3\,(Y -1)$.  The inferred dust masses are
listed in Table\,\ref{tab:dustparameters}. The inner dust shell
(observed with {\em Spitzer} IRS) contains only about one percent to
the total mass in both objects, but contributes roughly one sixths to
the mid-IR flux because of its higher temperature. The total dust mass
of \wrc\ is much higher than of \ka, reflecting its much stronger IR
flux.

%%%%%%%%%%%%%%%%%%%%%%%%%  FIG 15  %%%%%%%%%%%%%%%%%%%%%%%%%%%%%%%%
%-------------------------------------------------------------
% WITH DUSTY DATA
%
%_____________________________________________________________

\subsection{Molecular hydrogen diagnostics \label{sec:h2}} 

In the next two sections we discuss the emission lines in the {\em
Spitzer} IRS spectra that trace the gaseous material.

Emission from warm molecular hydrogen is routinely detected in
photo-dissociation regions (PDRs) (e.g.\ Parmar \etal\ \cite{par91}).
Models of H\,{\sc ii}/PDR regions predict a thin skin around the
ionized bubble, where hydrogen is mainly in atomic state (Kaufman
\etal\ \cite{kau06}). The PDR is adjacent to it. The PDR starts with a
transition region of warm ($\log{T}\sim 2\,..\,3$) molecular hydrogen
mixed with atomic H\,{\sc i}, which has $\approx 10$\,\% of the
spatial extend of the H\,{\sc ii} region. In the outer layers of the
PDR, the hydrogen becomes cooler ($\log{T}\sim 1\,..\,2$) and
predominantly molecular. The size of the H\,{\sc ii} region and,
consequently, the distance between the ionizing star and the zone
where the warm H$_2$ is located, depends on the number of ionizing
photons and on the electron density. For a star with
$\Phi_{\rm{i}}=10^{49}\,{\rm s}^{-1}$ and electron density $n_{\rm
e}=10$\,cm$^{-3}$, the PDR starts at $\msim\,10$\,pc from the ionizing
source (Kaufman \etal\ \cite{kau06}).  Ultracompact and compact
H\,{\sc ii} regions with radii $\lsim\,1$\,pc are observed around
new-born massive stars located in dense environments with $n_{\rm e}\msim
10^4$\,cm$^{-3}$ (Dopita \etal\ \cite{dop06}).

In the close vicinity of an evolved massive star the presence of
molecular hydrogen is very rare. Smith (\cite{smith02}) reported a
discovery of a ro-vibrational line
($\nu=1-0$)\,S(1)\,$\lambda$2.12\,\mim\ of H$_2$ in the Homunculus
nebula around $\eta$ Carinae. Here, we identify lines of
pure-rotational ($\nu=0$) transitions of molecular hydrogen at
$\lambda_{31} = 17.04$\,\mim\ [J=3-1] and $\lambda_{42} =
12.28$\,\mim\ [J=4-2] in the {\em Spitzer} spectra of \ka\ and \wrc\ (cf.\
Figs.\,\ref{fig:ka-irs}, \ref{fig:c-irs} and Table\,\ref{tab:irs}).
To our knowledge, this is the first detection of pure-rotational
transitions of molecular hydrogen in massive star nebulae.

A method to infer the excitation temperature and column density of
molecular hydrogen from the measured intensities of pure-rotational
lines was applied by Parmar \etal\ (\cite{par91}) to the Orion Bar.
When gas has sufficiently high density, collisions maintain the lowest
pure rotational levels of H$_2$ in thermal equilibrium (Burton \etal\
\cite{bht90}). Hence, the lowest rotational transitions of H$_2$
provide a thermometer for the warm gas. Rotational
transitions in the IRS band have small Einstein coefficients ($A_{42}$
= 2.76$\cdot$ 10$^{-9}$\,s$^{-1}$ and $A_{31}$ =
4.76$\cdot$10$^{-10}$\,s$^{-1}$, Turner \etal\ \cite{tur77}) and thus
are optically thin.

Since the interstellar extinction at $\lambda 17.04$\,\mim\ and $\lambda
12.28$\,\mim\ is similar, the ratio of intensities of these two optically thin
lines is
\begin{equation}
 \frac{I_{[3-1]}}{I_{[4-2]}}\,=\,\frac{N_{J=3}}{N_{J=4}}\frac{A_{31}}{A_{42}}
 \frac{\lambda_{42}}{\lambda_{31}}, 
\label{eq:rat}
\end{equation}
where $N_{J}$ is the column density of H$_2$ in level $J$.  Using the
Boltzmann equation for the ratio of the column densities, one can express the
temperature as a function of the observed line intensities:
\begin{equation}
kT = \left( E_{J=4}-E_{J=3}\right )/
{\ln{\left( {\cal A}\frac{I_{[3-1]}}{I_{[4-2]}} \right)}},
\label{eq:tex}
\end{equation}   
where $E_{J}$ is energy of upper level; for the two lines considered here,
$E_{J=4}/k$ = 1682\,K and $E_{J=3}/k$ = 1015\,K. The quantity $\cal{A}$
contains the constants for the considered line ratio,
\begin{equation}
{\cal A}\,=\,\frac{A_{42}}{A_{31}} \frac{\lambda_{31}}{\lambda_{42}} 
\frac{g_{J=4}}{g_{J=3}}. 
\label{eq:k}
\end{equation}
The statistical weights follow from $g_{\rm J}=(2J+1)(2I_n+1)$, 
where $I_n$ is the nuclear spin quantum number. $I_n$ is 0 for
even $J$ (para), and 1 for odd J (ortho), giving 
$g_{J=4}$ = 9 and $g_{J=3}$ = 21 for the upper levels of the considered 
transitions. Combining all constants yields $\cal A$ = 3.44 in our case. 

Inserting the line intensities from Table\,\ref{tab:irs} in
Eq.\,(\ref{eq:tex}), the temperature of warm molecular hydrogen gas in
the vicinity of \ka\ equals $\approx$460\,K. A higher temperature of the
H$_2$ gas is inferred from the IRS spectrum of \wrc, $T\approx 570$\,K.
It should be noted that the above results hold for a uniform density and
temperature, and therefore can be considered 
only as a rough estimates.
  
The column density of the molecular hydrogen can be estimated 
from the observed intensity in the H$_2$ lines. The line intensity
per steradian is given by
\begin{equation}
I(J)=h\nu_{JJ'} A_{JJ'} N_{(J)}/4\pi\,,
\label{eq:H2flux}
\end{equation}
where $N_{J}$ is the column density of molecules in the upper level $J$.
The column density of the total molecular hydrogen is inferred from the
Boltzmann equation, 
\begin{equation}
N({\rm H}_2)=N_{J}\,Z(T) \exp{(E_{J}/kT)}/g_J\,,
\label{eq:H2Boltzmann}
\end{equation}
where $Z(T)$ is the partition function (Herbst \etal\ (\cite{her96}):
\begin{equation}
Z(T) = 0.0247\ T \left[ 1 - \exp\frac{-6000{\rm K}}{T} \right]^{-1}.
\end{equation} 
The interstellar extinction is estimated for our program stars from
fitting the SED with stellar atmosphere models (see
Table\,\ref{tab:par}).  For the given wavelengths of the molecular
emission lines and the previously determined $E_{B\,-\,V}$, the
extinction weakens the intensity from the program stars by a factor
$10^{0.4A_\lambda} \approx 4.5$.

The de-reddened line intensities now enter Eq.\,(\ref{eq:H2flux}), and
the obtained number densities of the corresponding upper levels are
inserted into the Boltzmann Eq.\,(\ref{eq:H2Boltzmann}) yielding the
total H$_2$ column density. The resulting column density of warm
molecular hydrogen in the vicinity of \ka\ is $N({\rm H}_2)\approx
7\cdot 10^{20}$\,cm$^{-2}$, and in the vicinity \wrc\ $N({\rm
H}_2)\approx 6 \cdot 10^{20}$\,cm$^{-2}$.

We can estimate the thickness of the warm PDR zone. In the layer of
warm molecular hydrogen, the H\,{\sc ii}/PDR models predict roughly
the same number density for hydrogen in atomic form and for H$_2$
molecules (Kaufman
\etal\ \cite{kau06}). Taking thus the H$_2$ value for the H column
density, and $n_{\rm H}$ = 10$^4$ ... 10$^5$, the zone of warm
molecular hydrogen has only a width of 10$^{-3}$ ... 10$^{-2}$\,pc.

Furthermore, the information on the column density can be used to
derive the total mass of molecular hydrogen that is contained in the
column defined by the aperture of the {\em Spitzer} IRS
instrument. The size of the spectrograph slit corresponds to an area
of $A_{\rm slit}$ =0.18\,pc $\times$ 0.43\,pc = $7.6\cdot
10^{35}\,{\rm cm^2}$. The mass of H$_2$, $M_{\rm H_2} = 2\ m_{\rm H}\
N({\rm H}_2)\ A_{\rm slit}$, results as $0.8\,\Msun$ for \ka\ and
$0.7\,\Msun$ for \wrc.

Since the spectral energy distribution of the ionizing source is
known, we can constrain the distance of the PDR from the central star
using the combined H\,{\sc ii} region/PDR models presented by Kaufman
\etal\ (\cite{kau06}). For specified gas-phase elemental abundances
and grain properties, the parameters of a  model are the density of
H nuclei, $n_{\rm H}$, and the incident ``FUV'' flux, $G_0\propto
L_{\rm FUV}/r_{\rm PDR}^2$, where $G_0$ is expressed in units of
$1.6\cdot 10^{-3}$\,erg s$^{-1}$ cm$^{-2}$.  ``Far ultra violet''
(FUV) means energies above 6\,eV, but below the Lyman edge (13.6\,eV),
and $r_{\rm PDR}$ denotes the distance between the ionizing source and
the PDR.  Using the emergent flux from our atmosphere models, we
estimate $G_0 \times (r_{\rm PDR}/10{\rm pc})^{2} \approx$350 and
$\approx$210 for \ka\ and \wrc, respectively.

Kaufman \etal\ (\cite{kau06}) calculated intensities of the pure
rotational H$_2$ lines S(1) and S(2) as function of the hydrogen
density and the radiation-field parameter $G_0$ (see their Figs.\,5 and
6). Now we enter these diagrams with the measured S(1) and S(2) line
intensities for our program stars (Table\,\ref{tab:irs}). Because the 
numbers are very similar for both stars, we do not  
distinguish between them in the following order-of-magnitude estimates. 

The observed, line intensities are only reproduced for high density
and high radiation flux. Intensities of both H$_2$ lines, being
mutually consistent within a factor of two, allow a stripe in the
$n_{\rm H}$-$G_0$ parameter plane between $n_{\rm H}$ = 10$^4$\,
cm$^{-3}$, $G_0$ = 10$^5$ and $n_{\rm H}$ = 10$^5$\, cm$^{-3}$, $G_0$
= 10$^3$ (higher densities are not plausible and not covered by the
Kaufman \etal\ models).  By comparison with the $G_0 \times (r_{\rm
PDR}/10{\rm pc})^{2}$ values for the stellar radiation deduced above,
the models require a distance between star and PDR in the range 
0.6\,pc\,...\,6 \,pc.

\subsection{Parameters of the H\,{\sc ii} regions ionized by \ka\ and 
\wrc  
\label{sec:hydromass}}

Recently, Simpson \etal\ (\cite{sim07}) presented {\em Spitzer} IRS
(10-38\,\mim ) spectra obtained at 38 positions in the GC.
 The position 11 -- ``the Bubble Rim'' -- is $7\farcm 1$  away
from \ka, while the position 22 -- ``the Sickle Handle'' -- is only
$2\farcm 0$ away from \wrc.  

In the {\em Spitzer} SH IRS range, the line ratios $[$Ne\,{\sc
iii}$]$15.5\,\mim /$[$Ne\,{\sc ii}$]12.8$\,\mim\ and $[$S\,{\sc
iv}$]10.5$\,\mim /$[$S\,{\sc iii}$]18.7$\,\mim\ are indicative of the
excitation in the nebula.  In the IRS spectrum of
\wrc\ these ratios (see Table\,\ref{tab:irs}) are more than 
two times higher than those measured by Simpson \etal\ only $2\farcm
0$ away in the Sickle Handle, or anywhere else in their fields (see
their Table\,3).  Additionally, the continuum flux in the IRS spectrum
of \wrc\ is a few times higher compared to the observation of the
Sickle Handle.  This is in agreement with the IRAC and MSX images of
the field around \wrc\ shown in Figs.\,\ref{fig:irac-c} and
\ref{fig:msx-c}.  {\changed Also, Lang \etal\ (\cite{lang97}) comment on the
unusually high and unexplained (if only the Quintuplet cluster
ionizing stars are accounted for) ratio of H92$\alpha$ (8.31\,GHz) and
H115$\beta$ (8.43\,GHz) radio recombination lines in the southwest
Sickle ($l$$=$0.17,\,$b$$=$-0.40) where
\wrc\ is located. \wrc\ produces $\log(\Phi)= 50.12$ [s$^{-1}$]
ionizing photons, while the entire Quintuplet cluster produces
$\log(\Phi)= 50.5\,...\,50.9$ [s$^{-1}$] (Figer \etal\ \cite{Fig99}). 
\wrc\ is located at 3.5\,pc projected distance from the  Quintuplet,   
therefore the ionizing flux from \wrc\ dominates in a region of at
least 1\,pc around the star considering only geometrical dilution.  
We conclude that \wrc\ is the principle ionizing source
of the surrounding localized nebula that is located in the Sickle
Handle region and is probed by our {\em Spitzer} IRS spectrum.}

Similarly, the excitation ratios measured from our IRS spectra of
\ka\ are higher than those measured by Simpson \etal\ at the
Bubble Rim location. From the IRAC and MSX images of the field with
\ka\ (Figs.\,\ref{fig:irac-ka} and \ref{fig:msx-ka}) it is evident
that the nebula is centered on this WN9 type star. Given the high
temperature and luminosity of \ka\ it is  safe to conclude
that this star powers the IR emission from the surrounding H\,{\sc ii}
region.

To estimate the physical conditions in the H\,{\sc ii} regions around \ka\
and \wrc\ we use the dusty photoionization models calculated by Dopita
\etal\ (\cite{dop06}). Comparing the observed spectra with the model spectra 
from Dopita \etal, we roughly estimate the presure of ionized gas to
$\log{P/k\,\msim\,8}$\,[cm$^{-3}$\,K]. This is in agreement with the
location of \ka\ and \wrc\ in the  
\mbox{
$([$Ne\,{\sc iii}$]/[$S\,{\sc iii}$]$--$[$Ne\,{\sc iii}$]/[$Ne\,{\sc ii}$])$} 
and 
\mbox{
$([$Si\,{\sc iv}$]/[$S\,{\sc iii}$]$--$[$Ne\,{\sc iii}$]/[$Ne\,{\sc
ii}$])$} model diagrams.  Based on the available IRAC and MSX images and the
H$_2$ measurements, the size of the H\,{\sc ii} region around \ka\ is
not larger than 0.6\,...\,6\,pc.  According to the relationship
between the radius of the H\,{\sc ii} bubble and the pressure in the
ISM (Dopita
\etal\ \cite{dop05}), a bubble of such size around evolved stars is
only possible when the ISM pressure is large,
$\log{P/k\msim 8}$\,[cm$^{-3}$\,K]. This is significantly larger than
the ISM pressure in the GC on average (see e.g. Simpson \etal\
2007). Our measurements of the temperature and the density of the
H$_2$ (see Sect.\,\ref{sec:h2}) yielded $T_{\rm H2}$ $\approx$$500$\,K
and $n_{\rm H2}$ $\approx$$10^5$\,cm$^{-3}$, resulting in
$\log{P/k}\msim 7$\,[cm$^{-3}$\,K] in the PDR, where H$_2$ most likely
resides.

According to the diameter-density relationship from Dopita \etal\
(2006) the density in the H\,{\sc ii} region is $n_{\rm H}$
$\approx$$10^4$\,cm$^{-3}$ around \wrc\ and somewhat lower around the
cooler star \ka.  A lower limit to the mass of ionized gas is obtained
from $M_{\rm H\,II}=2A_{\rm slit}R_{\rm s}\,m_{\rm H}\,n_{\rm H}$,
where $A_{\rm slit}$ is the area of the {\em Spitzer} IRS slit, and
$R_{\rm s}$ is the Str\"omgen radius. The lower limit to the mass of
photoionized gas around \ka\ is $\approx$5\,$\Msun$ and around \wrc\
$\approx$10\,$\Msun$.

%%%%%%%%%%%%%%%%%%%%%%%%%%%%%%%%%%%%%%%%%%%%%%%%%%%%%%%%%%%%%%%%%%%%%%%%%%%%%
%--------------------  Section 5 Discussion ---------------------------------
%%%%%%%%%%%%%%%%%%%%%%%%%%%%%%%%%%%%%%%%%%%%%%%%%%%%%%%%%%%%%%%%%%%%%%%%%%%%%

\section{Discussion \label{seq:diss}}

\subsection{Stellar mass and evolution of \ka\ and \wrc}

Figer \etal\ (\cite{fig98}) presented evolutionary tracks for very
massive stars based on the code by Langer \etal\
(\cite{lan94}). According to these tracks, the initial stellar mass is
$150\,\lsim\,M_{\rm i}/\Msun\,\lsim\,200$ for \ka\ and
$100\,\lsim\,M_{\rm i}/\Msun\,\lsim\,150$ for \wrc. Thus, both our
program stars were initially among the most massive stars of the
Galaxy.

%%%%%%%%%%%%%%%%%%%%%%%%%  FIG 16 %%%%%%%%%%%%%%%%%%%%%%%%%%%%%%%%
%-------------------------------------------------------------
\begin{figure}
\centering
\includegraphics[width=\columnwidth]{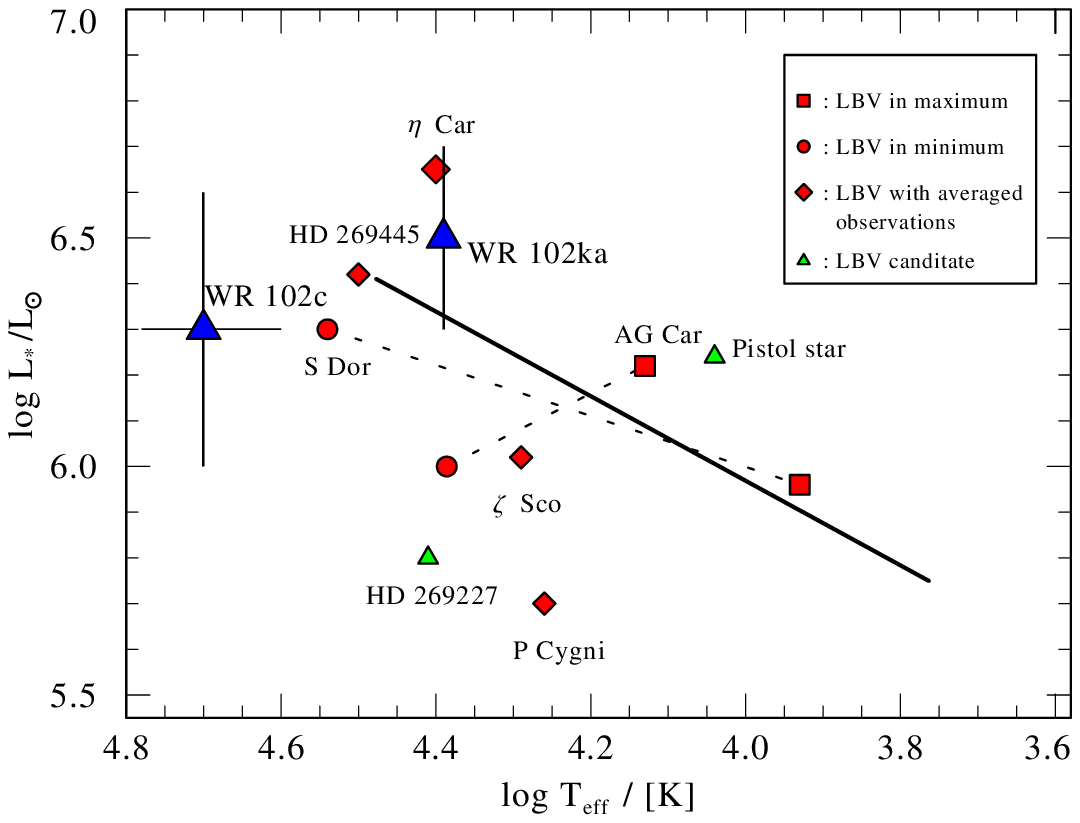}
\caption{
Hertzsprung-Russel diagram with the location of some known LBVs, LBV
candidates and our program stars. The thick line indicates the
Humphreys-Davidson limit (according Eq.\,(2) in Lamers \& Fitzpatrick
\cite{lam88}). The diamonds marks LBVs with time averaged
observations, squares shows the location of the LBV at the maximum,
circles at the minimum. The small triangles represent LBV candidates,
while the big triangles indicate the locations of \ka\ and \wrc. 
Except of \ka\ and \wrc, stellar parameters are from 
Figer \etal\ (\cite{fig98}), van Genderen (\cite{vg01}), 
Najarro  (\cite{naj06}), Groh \etal\ (\cite{gro06}).}
\label{fig:lbvhrd}
\end{figure}
%
%_____________________________________________________________

Our analysis of \ka\ yields an unconventionally high luminosity at a
relatively low stellar temperature (cf. Table\,\ref{tab:par}).  In the
HR-diagram (see Fig.\,\ref{fig:lbvhrd}) \ka\ is located above the
Humphreys-Davidson limit, in the region populated by the LBV
stars. \ka\ is of spectral type Ofpe/WN9. This class of objects
is often considered as either LBV candidates or LBVs in quiescence
(Crowther \etal\ \cite{cr95}, Morris \etal\ \cite{M1996}).  From the
analysis of the K-band spectrum, the surface mass fraction of hydrogen
in \ka\ is $0.2$. This is lower than the hydrogen mass fraction
0.3\,...\,0.4 found in known LBVs (Stothers
\& Chin, 2000).  The complimentary helium mass fraction in \ka\ 
is significantly higher than found for e.g. the Pistol star ($Y_{\rm
surf}\lsim 0.4$). This indicates a more advanced evolutionary stage of
the former, compared to ``normal'' LBV stars.  Interestingly, Maeder
\etal\ (\cite{maed08}) discussed possible filiations of Pop.\,I
massive stars. They suggest that stars with initial masses $M_{\rm i}
>90\,M_\odot$ do not pass through an LBV stage, but have high enough
mass loss to get rid of their envelopes on the main
sequence. According to this scenario, it is possible that \ka\ has not
been a classical LBV, but evolved directly from the Of to the WNL stage.

In the HR-diagram \wrc\ is located in the same region as the
outstandingly luminous WN6(h)\,...\,WN8(h) type stars WR\,22, WR\,24,
and WR\,25 (Hamann \etal\ \cite{wrh06}). An
orbital solution for the eclipsing binary WR\,22 yields $M_*\approx
70\,M_\odot$ (Rauw \etal\ \cite{rauw96}) in agreement with an
elaborate analysis based on the theory of optically thick winds
(Gr{\"a}fener \& Hamann \cite{gr08}).

A crude estimate of the present stellar mass of \wrc\ can be obtained
by using the scaling between $\vinf$ and the escape velocity
$\varv_{\rm{esc}}$ known for WN stars, 
$\vinf\,\approx\,1.5\,...\,4\,\varv_{\rm{esc}}$ 
(Lamers \& Cassinelli \cite{lam99}).  The escape
velocity is given by 
\mbox{$\varv_{\rm esc}\,=(2{\rm G}M_{\rm eff}/R_\ast)^{0.5}$}, 
where \mbox{$M_{\rm eff}=M_\ast(1-\Gamma)$}, and $\Gamma$ is the
Eddington factor. Using stellar parameters and abundances of \wrc\
from Table\,\ref{tab:par} we estimate the present stellar mass to
45\,...\,55\,\Msun.

Because the scaling between $\vinf$ and the $\varv_{\rm{esc}}$ is not
established for stars located above the Humphreys-Davidson limit, the
same method cannot be applied to constrain the present mass of \ka.

A vast amount of chemically enriched material has been lost during
recent stellar evolution of \ka\ and \wrc, and has contributed to the
nebulae around these stars.  Interestingly, the evolutionary tracks
(Meynet \etal\ \cite{mae03}) predict that the carbon to oxygen
mass-ratio on the surface is larger than unity already in the WN phase
for stars with $M_{\rm i} \msim 85\,\Msun$. Thus one may speculate
that the PAH features observed in the spectra of our program stars and
elsewhere in the GC reflect the carbon enrichment of the
ISM by the stellar winds from initially very massive stars.

Neither \ka\ nor \wrc\ belong to the central parts of stellar
clusters. They may have been either dynamically ejected from parental
clusters, or formed in isolation.  Conspicuously, \ka\ is located at
the Bubble Rim, where the ISM could be pressurized by the expanding hot
bubble. \wrc\ is located in the Sickle nebula at the edge of a dense
molecular cloud that is ionized by the Quintuplet cluster (Simpson
\etal\ \cite{sim97}).  The Quintuplet cluster is  3-5\,Myr old and 
contains a large population of WC-type stars that are older than
\wrc. {\changed Moreover, the radial velocity of the Quintuplet
cluster, $130$\,km\,s$^{-1}$ (Figer \etal\ \cite{Fig99}), is quite
different from the average radial velocity of the Sickle nebula, $\sim
35$\,km\,s$^{-1}$ (Lang \etal\ \cite{lang97}). Hence, most likely
formation of \wrc\ is not linked to the Quituplet cluster.}

%%%%%%%%%%%%%%%%%%%%%%%%%  FIG 17 %%%%%%%%%%%%%%%%%%%%%%%%%%%%%%%%
%%-------------------------------------------------------------
\begin{figure}
\centering
\includegraphics[width=\columnwidth]{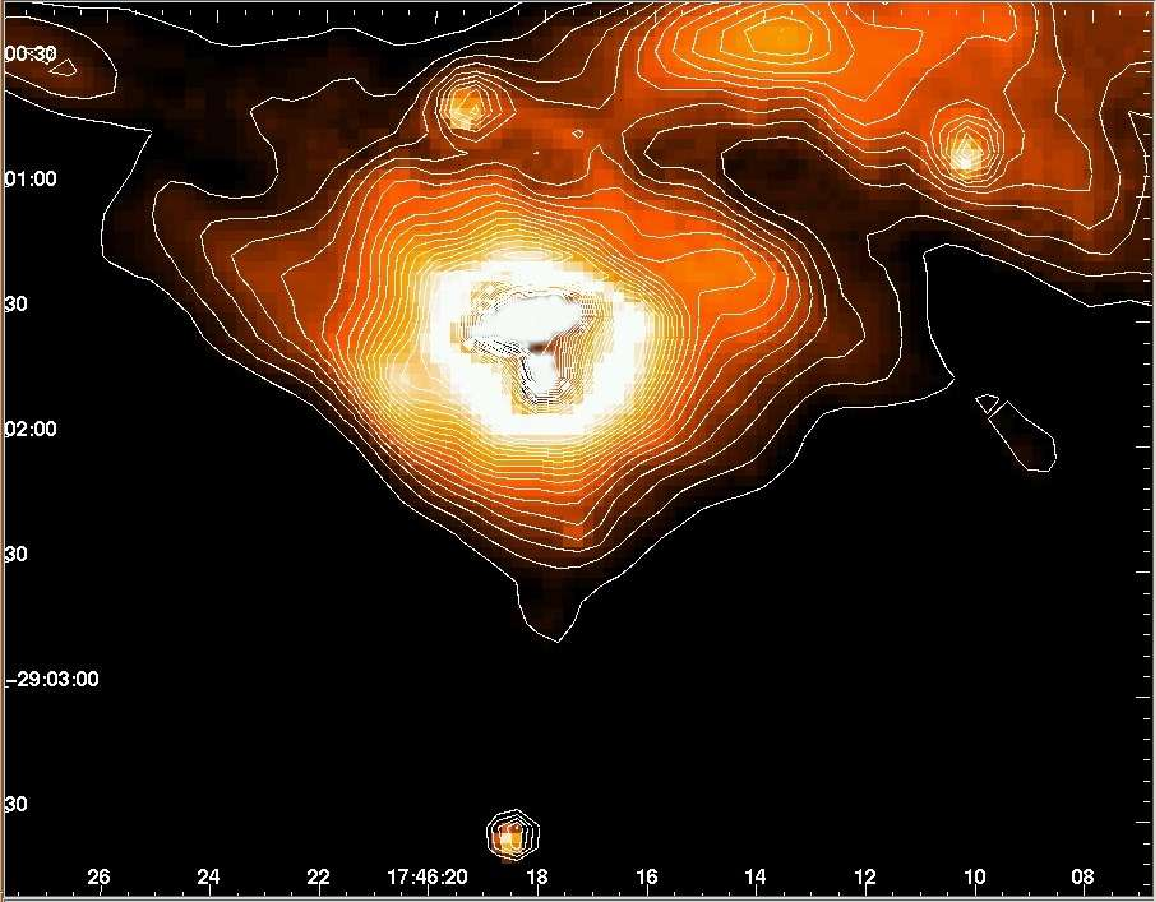}
\caption{{\em Spitzer} MIPS 24\,\mim\  image  of \ka\ and its peony-shaped
nebula. 
The spatial resolution is $2\farcs 5$. The nebula has a radius of
about $50$\,\arcsec. The colors are chosen such that the background
appears black. The image is shown on linear scale, as well as the
contours. Two central parts of the image are saturated, and are shown
in white for cosmetic reasons. North is up, east to the left.}
\label{fig:mipska}
\end{figure}
%
%%_____________________________________________________________
%

\subsection{Nature of the nebula around \ka}

Recently, the {\em Spitzer} MIPS survey of the GC at 24\,\mim\ became
publically available. The 24\,\mim\ image of \ka\ and its nebula is
shown in Fig.\,\ref{fig:mipska}.  It confirms what was already obvious
from the less sensitive MSX observation -- the presence of a compact
dusty nebula centered on \ka. The MIPS image resolves its roughly
spherical shape, resembling a Peony flower.  The radius of the ``peony
nebula'' in 24\,\mim\ image is about 50\,\arcsec ($\approx$1.5\,pc),
i.e.  nearly two times larger than the size of the unresolved image in
the MSX E-band.  This is because the MIPS camera is $\msim$$10^4$
times more sensitive.  Unfortunately, the MIPS image is saturated in
central parts, and therefore fluxes cannot be extracted for modeling
purposes. Intriguingly, the two brightest, saturated parts of the
``peony nebula'' coincide with the point-like central star and with
the fuzzy, feather-like feature north of the star, visible in the
8\,\mim\ IRAC image (right panel of Fig.\,\ref{fig:irac-ka}). This
spatial coincidence is interesting, because the IRAC 8\,\mim\ images
trace the emission from PAH, while the MIPS 24\,\mim\ image shows the
emission from small grains. For our {\em Spitzer} observation the slit
was centered on \ka, and therefore our spectrum samples the central
part of the nebula.

Unfortunately, the spectrum covers only the short-wavelength range and
does not allow a detailed study of the chemical composition and the
dynamics of the \ka\ nebula. The available part of the nebula spectrum
is similar to the spectra of known LBV nebulae (cf.\ Voors \etal\
\cite{voors}). As discussed above, \ka\ is either a 
post-LBV star or a star that suffered high mass loss already on the
main sequence. It is natural to suggest that the photoionized
peony-shaped nebula contains stellar material that was lost by \ka\
during LBV-type eruptions and/or its strong stellar wind.

\section{Summary}

Table\,\ref{tab:mass} compiles some of the results discussed in the previous 
sections.   

%---------------------------------------------------------
\begin{table}
\caption{Masses and radii. All numbers must be considered as rough estimates.}
\label{tab:mass}
\centering
\begin{tabular}{lcc}
\hline
\hline
               & \ka\  & \wrc\ \\ \hline 
 Star:          &     &         \\    
Initial mass    & 150\,...\,200\,$M_\odot$ & 100\,...\,150\,$M_\odot$ \\
Current mass    &      ?                   & 45\,...\,55\,$M_\odot$ \\  \hline
H\,{\sc ii} region:&    &  \\
Ionized gas     & $>5\,M_\odot$   &  $>10\,M_\odot$  \\   
Radius          & 0.6\,...\,6\,pc & 0.6\,...\,6\,pc \\  \hline
PDR:            &             &         \\
H$_2$ mass      &  $>0.8\,M_\odot$   & $>0.7\,M_\odot$  \\ \hline
Dust shell:     &                    &                   \\
Warm dust mass  & 0.005\,$M_\odot$   &  0.08\,$M_\odot$  \\
Inner radius    & 0.06\,pc           & 0.07\,pc  \\ 
Outer radius    & 0.4\,pc            & 0.4\,pc   \\  
 \hline    
\end{tabular}
\end{table}
%------------------------------------------------------------

\noindent
1) \wrc\ and \ka\ are among the most luminous and initially most
massive stars in the Galaxy. 

\medskip\noindent
2) In the HR diagram, \wrc\ shares  its location with the overluminous, 
very massive WN-type stars WR\,22, WR\,24, and WR\,25.

\medskip\noindent
3) \ka\ is located above the Humphreys-Davidson limit in the HR
diagram.  Its initial mass plausibly was in excess of $150\,M_\odot$.
The K-band spectrum of \ka\ shows indications that the star is a
WNE+Of/WN9 binary. A hypothetical WNE companion has a higher effective
temperature but lower luminosity than the primary.

\medskip\noindent
4) The {\em Spitzer} IRS spectra of \ka\ and \wrc\ are dominated by
emission of a dusty nebulae powered by the stellar radiation of their
respective central stars.  Based on the modeling of the spectral energy
distribution, the inner radius of the circumstellar dust shells is
$\approx$$10^3\,R_\ast$. This is the first detection of dust so close to
the surface of a WN-type star.

\medskip\noindent
5) The lines of pure-rotational transitions of molecular hydrogen are
detected in the nebular spectra of \ka\ and \wrc. To our knowledge,
this is the first detection of pure-rotational transitions of warm
H$_2$ in the spectra of nebulae around evolved massive stars. The mass
of the warm molecular hydrogen is about one solar mass in each
nebulae.

\medskip\noindent
6) Assuming that H${_2}$ lines originate in the PDRs, the radii of the
H\,{\sc ii} regions around \ka\ and \wrc\ are in the range
0.6\,...\,6\,pc.  These radii are significantly smaller than normally
expected for evolved hot massive stars, and probably reflect the high
density and pressure in the respective H\,{\sc ii} regions and the
ISM.

\medskip\noindent
7) \wrc\ is the {\changed dominant ionizing source of the rather
compact H\,{\sc ii} region located in the Sickle Handle.}  This
demonstrates the importance of individual massive stars scattered in
the field as excitation sources in the GC.

\medskip\noindent
8) The nebula powered by \ka\ is resolved in the MIPS 24\,\mim\ image.
This ``Peony nebula'' contains dust and warm molecular hydrogen.  We
suggest that the formation of the nebula is a result of strong recent
mass loss by \ka.

\begin{acknowledgements}

This work is based on observations obtained with the {\em Spitzer}
Space Telescope, which is operated by the Jet Propulsion Laboratory,
California Institute of Technology under a contract with NASA.  This
work has extensively used the NASA/IPAC Infrared Science Archive, the
NASA's Astrophysics Data System, and the SIMBAD database, operated at
CDS, Strasbourg, France. {\changed The comments of an anonymous 
referee helped to improve the clarity of the paper.} The
authors thank Nicole Homeier for providing the K-band spectrum of \ka,
and M.W.\,Pound and M.\,Wolfire for their advices regarding the PDR
Toolbox (http://dustem.astro.umd.edu).

\end{acknowledgements}

%\end{document} 


\begin{thebibliography}{}

\bibitem[1990]{bht90} 
Burton, M., Hollenbach, D.J.,  \& Tielens, A.G.G.M. 1990, \apj, 365, 620

\bibitem[2005]{lbv2005} 
Clark, J.\,S., Larionov, V.\,M., \& Arkharov, A. 2005, \aap, 435, 239 

\bibitem[2001]{cg01} 
Cohen, M., \& Green, A.J. 2001, \mnras, 325, 531

\bibitem[2000]{cot00} 
Cotera, A.S., Simpson, J.P., Erickson, E.F., \etal\ 2000, ApJS, 129, 123

\bibitem[1995]{cr95} 
Crowther, P.A., Hillier,D.J., \& Smith, L.J. 1995, \aap, 293, 172

\bibitem[2006]{wc9} 	
Crowther, P.A., Morris, P.W., \& Smith, J.D. 2006, \apj, 636, 1033

\bibitem[2005]{dop05} 
Dopita, M.A., Groves, B.A., Fischera, J., \etal\ 2005, ApJ, 619, 755

\bibitem[2006]{dop06} 
Dopita, M.A., Fischera, J., Crowley, O., \etal\ 2006, \apj, 639, 788

\bibitem[2004]{dr04}
Draine, B.T., 2004, in Origin and Evolution of the Elements, 
Carnegie Observatories Centennial Symposia, ed. A. McWilliam \& M. Rauch, 317

\bibitem[1984]{dr84} 
Draine, B.T., \& Lee, H.M. 1984, \apj, 285, 89 

\bibitem[2004]{irac} 
Fazio G.G., Hora, J.L., Allen, L.E., \etal\ 2004, ApJ Suppl., 154, 10

\bibitem[1998]{fig98}
Figer, D.F., Najarro, F., Morris, M., \etal\ 1998, \apj, 506, 384

\bibitem[1999a]{Fig99} 
Figer, D.F., McLean, I.S., \& Morris, M. 1999a,  \apj, 514, 202

\bibitem[1999b]{quin99}
Figer, D.F., Kim, S.S., Morris, M., \etal\ 1999b, \apj, 525, 750

%\bibitem[1999c]{figpn99}
%Figer, D.F., Morris, M., Geballe, T.R., \etal\ 1999, \apj, 525, 759

\bibitem[2003]{fr03}
Freyer, T., Hensler, G., \& Yorke, H. W. 2003, \apj, 594, 888

\bibitem[2006]{ful06}
Fullerton, A.W., Massa, D.L., \& Prinja, R.K. 2006, ApJ, 637, 1025

\bibitem[2008]{gr08}
Gr{\"a}fener, G., \& Hamann, W.-R. 2008, A\&A, 482, 945

\bibitem[2006]{gro06}
Groh, J.H., Damineli, A., Teodoro, M., \& Barbosa, C.L. 2006, A\&A, 457, 591

\bibitem[2007]{had07}
Hadfield, L.J., van Dyk, S.D., Morris, P.W., \etal\ 2007, \mnras, 376, 248

\bibitem[1991]{ham91} 
Hamann, W.-R., D{\"u}nnebeil, G., Koesterke, L., Schmutz, W., 
\& Wessolowski, U. 1991, A\&A, 249, 443 

\bibitem[1998]{hk98} 
Hamann, W.-R., \& Koesterke, L. 1998, \aap, 333, 251

\bibitem[2004]{wrh04} 
Hamann, W.-R., \& Gr{\"a}fener G. 2004, \aap, 427, 697
  
\bibitem[2006]{wrh06} 
Hamann, W.-R., Gr{\"a}fener G., \& Liermann A. 2006, \aap, 457, 1015

\bibitem[1996]{her96}
Herbst, T.M., Beckwith, S.V.W., Glindemann, A., \etal\ 1996, \apj, 111, 2403

\bibitem[2003]{H03} 
Homeier, N.L., Blum, R.D., Pasquali, A., Conti, P.S., \& Damineli, A. 2003,  
\aap, 408, 153

\bibitem[2004]{houck04} 
Houck, J.R., Roellig, T.L., van Cleve, J., \etal\ 2004, ApJS, 154, 18

\bibitem[1997]{E1997} 
Ivezi\'c, Z., \& Elitzur, M. 1997, \mnras, 287, 799

\bibitem[2006]{kau06}
Kaufman, M.J., Wolfire, M.G., \& Hollenbach, D.J. 2006, \apj, 644, 283

\bibitem[1995]{krab95}
Krabbe, A., Genzel, R., Eckart, A., \etal\ 1995, ApJ, 447, L95

\bibitem[1988]{lam88}
Lamers, H.J.G.L.M., \& Fitzpatrick, E.L. 1988, \apj, 324, 279

\bibitem[1999]{lam99}
Lamers, H.J.G.L.M., \& Cassinelli, J.P. 1999, Introduction to 
Stellar Winds, ed. H.J.G.L.M. Lamers, \& 
J.P. Cassinelli, ISBN 0521593980 (Cambridge, UK: Cambridge University 
Press), 49

\bibitem[1997]{lang97}
Lang, C.C., Goss, W.M., \& Wood, D.O.S. 1997, ApJ, 474, 275

\bibitem[1994]{lan94}
Langer, N., Hamann, W.-R., Lennon, M., Najarro, F., Puls, J., \&
Pauldrach, A. 1994, \aap, 290, 819

\bibitem[2005]{Li05} 
Li, A. 2005, AIPC, 761, 123
 
\bibitem[2008]{maed08}
Maeder, A., Meynet, G., Ekstr{\"o}m, S., Hirschi, R., \& Georgy, C. 2008, 
astro-ph/0801.4712 

\bibitem[2003]{mae03}
Meynet, G., \& Maeder, A., 2003, \aap, 404, 975

\bibitem[2007]{mar07}	
Martins, F., Genzel, R., Hillier, D.J., \etal\ 2007, \aap, 468, 233

%\bibitem[2008]{mar08}	
%Martins, F., Hillier, D.J., Paumard, T., Eisenhauer, F., Ott, T., \& Genzel, R.
% 2008,  \aap, 478, 219

\bibitem[1977]{m1977}
Mathis, J.S., Rumpl, W., \& Nordsieck, K.H. 1977, \apj, 217, 425

\bibitem[2001]{m2001}
Moneti, A., Stolovy, S., Bommart, J.A.D.L, \etal\ 2001, \aap, 366, 106

\bibitem[1996]{M1996}
Morris, P.W., Eenens, P.R.J, Hanson, M.M., \etal\ 1996, \apj, 470, 597

\bibitem[2006]{naj06}
Najarro, F. 2006, Journal of Physics Conference Series, 54, 224

\bibitem[2007]{osk07}
Oskinova, L.M., Hamann, W.-R., \& Feldmeier, A. 2007, A\&A, 476, 1331

\bibitem[1991]{par91}
Parmar, P.S., Lacy, J.H., \& Achtermann, J.M. 1991, \apj, 372, 25

\bibitem[2001]{msx01}
Price, S.D., Egan, M.P., Carey, S.J., Mizuno, D.R., \& Kuchar, T.A. 2001, 
AJ, 121, 2819

\bibitem[1996]{rauw96}
Rauw, G., Vreux, J.-M., Gosset, E., \etal\ 1996, A\&A, 306, 771 

\bibitem[2005]{iracm}
Reach, W., Megeath, S.T., Cohen, M., \etal\ 2005, PASP, 117, 978 

\bibitem[1993]{r93}
Reid, M.J., ARA\&A, 31, 345

\bibitem[2008]{ram08}
Ram{\'i}rez, S.V., Arendt, R.G., Sellgren, K., \etal\ 2008, ApJS, 175, 147

\bibitem[2001]{rf01}
Rodriguez-Fern{\'a}ndez, N.J., Martín-Pintado, J., \& de Vicente, P. 2001, 
A\&A, 377, 631

%\bibitem[1996]{r96}
%Roelfsema, P., Cox, P., Tielens, A., \etal\ 1996, A\&A, 315, L289 

\bibitem[1999]{s1999}
Schultheis, M., Ganesh, S., Simon, G., \etal\ 1999, \aap, 349, L69

\bibitem[1998]{arch98} 
Serabyn, E., Shupe, D., \& Figer, D.F., 1998, Nature, 394, 448

\bibitem[1997]{sim97}
Simpson, J. P., Colgan, S.W.J., Cotera, A.S., \etal\ 1997, ApJ, 487, 689

\bibitem[2007]{sim07}
Simpson, J., Colgan, S.W.J., Cotera, A.S., \etal\ 2007, \apj, 670, 1115

\bibitem[2002]{smith02}
Smith, N. 2002, MNRAS, 337, 1252

\bibitem[2008]{smith08}
Smith, N., \& Conti, P.S. 2008, ApJ, 679, 1467

\bibitem[2006]{sto06} 
Stolovy, S., Ramirez, S., Arendt, R.G., \etal\ 2006, JPhCS, 54, 176 

\bibitem[2000]{sto2000}
Stothers, R.B., \& Chin, C.W. 2000, \apj, 540, 1041

\bibitem[1977]{tur77}
Turner, J., Kirby-Docken, K., \& Dalgarno, A. 1977, \apjs, 35, 281

\bibitem[2001]{vg01}
van Genderen, A.M. 2001, A\&A, 366, 508

%\bibitem[2005]{vm05} 
%van Marle, A.J., Langer, N., \& Garc{\'i}a-Segura, G. 2005, \aap, 444, 837

\bibitem[2000]{voors} 
Voors, R.H.M., Waters, L.B.F.M., de Koter, A., \etal\ 2000, \aap, 356, 501

%\bibitem[1997]{wil97} 
%Williams, P.M. 1997, Ap\&SS, 251, 321

\bibitem[1987]{yz87} 
Yusef-Zadeh, F., \& Morris, M. 1987, ApJ, 320, 545

%\bibitem[1998]{zub98}
%Zubko, V.G. 1998, \mnras, 295, 109   

\end{thebibliography}
\end{document}